\documentclass[conference]{IEEEtran}
\IEEEoverridecommandlockouts
\AtBeginDocument{%
  \providecommand\BibTeX{{%
    \normalfont B\kern-0.5em{\scshape i\kern-0.25em b}\kern-0.8em\TeX}}}

\usepackage{booktabs} % For formal tables
\usepackage{colortbl}
\usepackage{amsthm}
\usepackage{color,soul,xspace}
\usepackage{amsbsy}
\usepackage{multirow}

\usepackage[tight,footnotesize]{subfigure}
\usepackage{caption}
\usepackage{float}
\usepackage{graphicx}
\usepackage{caption}
\usepackage[tight,footnotesize]{subfigure}
\usepackage[linesnumbered, ruled]{algorithm2e}
\usepackage[noend]{algpseudocode}
\usepackage{mdframed}

\newcommand{\eat}[1]{}

\makeatother
\DeclareMathAlphabet\mathbfcal{OMS}{cmsy}{b}{n}
\newcommand{\todo}[1]{\sethlcolor{red}\hl{TODO: #1}}

\newcommand{\mourmeth}{\text{DSEN}}
\newcommand{\ourmeth}{$\mourmeth$\xspace}

\usepackage{tabularx,lipsum,environ,amsmath,amssymb}

\makeatletter
\newcommand{\problemtitle}[1]{\gdef\@problemtitle{#1}}% Store problem title
\newcommand{\probleminput}[1]{\gdef\@probleminput{#1}}% Store problem input
\newcommand{\problemquestion}[1]{\gdef\@problemquestion{#1}}% Store problem question
\NewEnviron{problem}{
  \problemtitle{}\probleminput{}\problemquestion{}% Default input is empty
  \BODY% Parse input
  \par\addvspace{.5\baselineskip}
  \noindent
  \begin{tabularx}{\textwidth}{@{\hspace{\parindent}} l X c}
    \multicolumn{2}{@{\hspace{\parindent}}l}{\@problemtitle} \\% Title
    \textbf{Input:} & \@probleminput \\% Input
    \textbf{Question:} & \@problemquestion% Question
  \end{tabularx}
  \par\addvspace{.5\baselineskip}
}
\makeatother
\begin{document}

\title{ 
A Large-scale Friend Suggestion Architecture
\\
\thanks{*Authors are equally contributed.}
}

\author{\IEEEauthorblockN{1\textsuperscript{st} Lin Zhang$^{*}$}
\IEEEauthorblockA{\textit{International Digital Economy Academy (IDEA)} \\
% \textit{IDEA.EDU}\\
Shenzhen, China \\
zhanglin@idea.edu.cn}
\and
\IEEEauthorblockN{2\textsuperscript{nd} Rui Li$^{*}$}
\IEEEauthorblockA{\textit{Harbin Institute of Technology} \\
Shenzhen, China \\
19s152089@stu.hit.edu.cn}
}

\maketitle
\begin{abstract}
Online social as an extension of traditional life plays an important role in our daily lives. 
Users often seek out new friends that have significant similarities such as interests and habits, motivating us to exploit such online information to suggest friends to users. In this work, we focus on friend suggestion in online game platforms because in-game social quality significantly correlates with player engagement, determining game experience. 
Unlike a typical recommendation system that depends on item-user interactions, in our setting, user-user interactions do not depend on each other. Meanwhile, user preferences change rapidly due to fast changing game environment. 
There has been little work on designing friend suggestion when facing these difficulties, and for the first time we aim to tackle this in large scale online games. 
Motivated by the fast changing online game environment, we formulate this problem as friend ranking by modeling the evolution of similarity among users, exploiting the long-term and short-term feature of users in games. 
Our experiments on large-scale game datasets with several million users demonstrate that our proposed model achieves superior performance over other competing baselines.
\end{abstract}

\section{Introduction }

With the rapid development of the World Wide Web, most of the human activities are migrated to the Internet,
such as %politics~\cite{AdamickKDD2005}, 
game~\cite{Zhang2017SIGIR} and social~\cite{KwakWWW2010}.  %
Online service providers often develop efficient toolboxes to improve our online experience by analyzing our activities. 
% One of the most successful practice is the recommendation systems~\cite{Tang2013Socialrecommendation} that analyzes the users' behaviors, such as viewing and purchasing, and then recommends items to users to fit the obtained profiles of individual users.
A popular practice is recommender systems~\cite{Tang2013Socialrecommendation}, which analyze user behavior like viewing and purchasing, and then recommend the items to users so as to fit the obtained profiles of the individual users. 
Apart from the need of items, we naturally desire to make friends because friendship constitutes a critical part of humanity. 
In recent years, the development of recommendation systems is mostly concentrated in user-item setting~\cite{BOBADILLA20131}, however, recommending strangers to make friends is also a desirable feature in many applications like players in games and dating applications. 
%  particular in the case that users are anonymous, 
\emph{How can we suggest new friends to users in an online setting when their interactions are sparse, mostly unavailable in reality?} In this work, we focus on online games and refer to this setting as friend suggestion~\cite{Maayankdd2010}.%

\begin{figure}[t]
    \centering
    \includegraphics[trim={0 0cm 4cm 12cm},clip,width=0.48\textwidth]{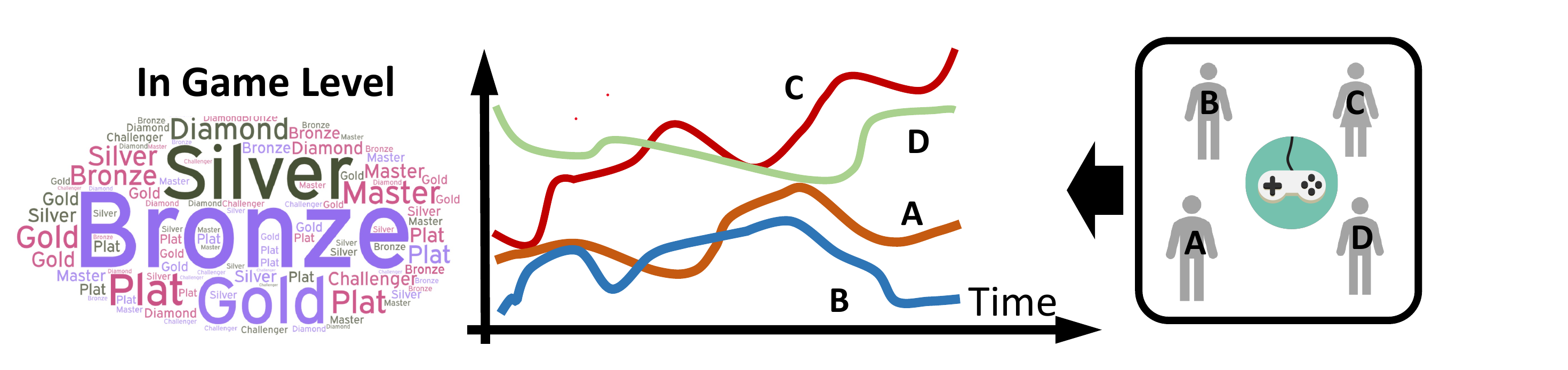}
    \caption{
Online games often provide fast updates to the social environment, and players tend to react quickly to changes in  games.
    For example, one common feature of users is their game level, which changes rapidly and so that the preference on choosing friends changes accordingly.  This phenomenon suggests that a good friend suggestion system should adapt to this fast changing preference as opposed to a static user preference in a user-item recommendation setting. }
    %\todo{lin:add more details}}
    \label{fig:intro_overview}
\end{figure}

From the theory of homophily in sociology~\cite{McPherson2001}, we know that people tend to associate with people that have similar behaviors to each other, and people that we friend with tend to behave like each other over time. This motivates us that the key for the friend suggestion task is to find people that are similar to each other, and therefore we need to address the problem of estimating the similarity between users.
To be able to estimate similarity, information about user profiles~\cite{Raadprofile2010,Peled2013} and behaviors~\cite{Chen2014,Nguyen2018} %Liu2012
has to be taken into account. 
It provides two different perspectives to model user preferences on decision making, including long-term and short-term intents. 
Users from online games generate abundant data consisting of both perspectives. 
Specifically, individual game players generate rich online activities from time to time, and many of them are collected in a sequential order that reflects user dynamics, capturing users' short-term changes.
By contrast, players' profiles provide a long-term perspective to describe players, since these contain slow change or fixed information such as gender and age, which often have a predominant influence on players' decisions. 
By fusing long-term and short-term information, we can obtain a more comprehensive representation of users and build a model on top of it to learn the similarity between users. 

One popular way to exploit the aforementioned short-term and long-term features is by sequential recommendation~\cite{Chen2019BehaviorST} from traditional e-commerce recommendation settings, leveraging sequences of viewing or purchasing items in online shopping platforms. 
One may argue that we can collect user-user interaction sequences and apply existing sequential recommendation models~\cite{2019Deep}.  % 2017Deep Chen2019kdd % Quadrana2018survey
However, the direct interaction behavior of users is often sparser than that of user-item interactions~\cite{Zeng2013InformationFI} because common online websites or mobile apps are often not designed for social interaction directly.
Beyond the sparseness in interaction,  users' behaviors in games may also be sparse due to the inactivity of large proportion of users, which is a typical phenomenon in games.
Particularly, interactions among users has significant difference to that of user-item setting.
On one hand, a user-item interaction sequence has consecutive dependencies over each other, reflecting a user's sequential decisions~\cite{ wang2019sequential}. % ji2019sequential
Although user-user interactions happened sequentially, 
the consecutive interactions are not necessarily depending on each other as those in a user-item interaction sequence, suggesting that
we can not simply treat them as a logical sequence and use these existing sequential recommendation models to address our problem.
On the other hand, a user's preference over items changes slowly, since the general taste of a user is often static with no or slow change over time~\cite{yakhchi2021learning}.
While this may not hold for players in games because of the fast updating environment in games, as shown in Fig.~\ref{fig:intro_overview}.
In contrast to a real-world human that takes years to develop his/her taste, the role of a player in the game changes rapidly to adapt to the surrounding, making his/her preferences about the decision evolves quickly rather than static. The question then arises: \textit{how can we deal with friend suggestion in game platforms where users' preferences change rapidly, along with non-logical and sparse user-user interactions? 
}

To address the aforementioned issue, we propose a novel end-to-end solution that leverages user behavior evolution with the proposed similarity evolution modeling, called Deep Similarity Evolutionary neural Network ~(\ourmeth). 
Specifically, we first learn a low-dimensional embedding from user features at each time point through a sequence sensitive model,
%by applying Gated Recurrent Units~\cite{chung2014empirical}, 
fusing the long-term information and the short-term information as the unified embedding. 
We then characterize the similarity between users over time by computing they similarity through an efficient generalized dot product model.
Next, we model the evolution of user preference over time using the obtained similarity along the time line implicitly,
% which is formalized by LSTM~\cite{2014Learning}, 
with the goal of inferring the current preference from the past. 

To sum up, we have the following contributions:\\
1, \textbf{Novelty.} We develop a novel end-to-end friend suggestion framework by integrating both long-term and short-term information. 
To the best of our knowledge, this can be the first work on this task without relying on interactions among users, and also the first application to online game scenarios. \\
2, \textbf{Efficiency.} We evaluate our model on large scale data generated from a popular online game with several millions of users, and show superior results over competing methods. \\

\section{Related work }

We now brief review the related work in the literature. We can categorize the related into the following topics.

\emph{Friend suggestion:}
% Most related work comes from the topic of friend suggestion, sharing the same scope with our work.
Existing friend suggestion methods often rely on mature social networks, assuming that users have solid and frequent connections with others, then learning features from this graph structure to measure the similarity between users, such as using mutual social influence~\cite{Zheng2017FriendRI}, using scholar interaction~\cite{XU2019}, using social network~\cite{SankarWWW2021}. % EpastoVLDB2015
% and cross marketing~\cite{LI2020}. 
%, such as  location-based mobile information~\cite{Chu2013}, web information~\cite{ADAMIC2003}, mutual social influence~\cite{Zheng2017FriendRI}, scholar interaction~\cite{XU2019}, social network~\cite{EpastoVLDB2015,SankarWWW2021} and cross marketing~\cite{LI2020}. 
In ~\cite{EpastoVLDB2015}, authors developed a social community-based model to deal with friend suggestions by analyzing users' co-occurrences in neighbors. 
In ~\cite{suggest}, they proposed a friend suggestion algorithm that uses a user's implicit social graph to generate a friend group. Their method apply to any interaction-based social network.
% \cite{}
Despite its effectiveness, such methods may not fit in our setting from two perspectives: One is that we may not have access to stable social networks, and another is that user-user interaction is sparse in games. Some players even intentionally avoid engaging with friends and familiarity, thus making the social network useless. 
% Among these, the Graph Neural Networks (GNN)~\cite{zhou2021graph} 
% gains more and more attention and has been applied to friend recommendation by delivering powerful embedding~\cite{SankarWWW2021}.
% In ~\cite{suggest}, they proposed a friend suggestion algorithm that uses a user's implicit social graph to generate a friend group. Their method apply to any interaction-based social network but do not utilize the user's action sequence and profile vector well which is the core of our method.
% do not utilize the user's action sequence and profile vector well which is the core of our method.
Meanwhile, Graph Neural Networks (GNNs)~\cite{zhou2021graph}  are gaining more and more attention and have been applied to friend recommendation by supplying powerful embeddings~\cite{SankarWWW2021}.
Despite showing promising results, these methods rely on clear interactions and well-defined graphs. 
Existing methods may not be applicable when no solid social network is available. 
% \todo{add more details}

\emph{Sequential recommendation system:}
One related topic is sequential recommendation, focusing on predicting the next user interaction with items~\cite{Guo2020www, Wang2015SIGIR}  by learning user interaction sequences in the history. 
Prior to the era of deep leaning, researchers have applied Markov chain theory to sequence modeling and make prediction on the chain~\cite{he2016fusing}. 
Such models assume that the next actions depend on the most up-to-date actions, therefore these fail to capture long-term dependencies. 
More recently, models based on Recurrent Neural Networks (RNN) have overcome the limits of Markov chain and gained significant progress~\cite{Yu2016SIGIR}. % Zhang2014AAAI
Beyond RNNs, its variants including Long Short-Term Memory (LSTM)~\cite{graves2014generating} and Gated Recurrent Unit (GRU)~\cite{Chung2014NIPS} achieve further improvement by addressing issues in conventional RNNs. 
% TransFormer and Bert are word embedding models based on dynamic representation. 
Beyond such recurrent-based models, recent advanced models like TransFormer and Bert show better ability to capture the user's short-term and long-term interests from a sequence of user-item interactions~\cite{self-rec, bert4rec}.
Besides, researchers employ GNNs  to address this problem~\cite{SRGNN}, % SURGE
converting the user's sequence data into a sequence graph.
% can make the original item selection conversion more flexible
% GNNs can capture the complex user preferences implicit in the sequence's behavior via ring structure. 
Yet, as these sequences are fundamentally different from ours, their interaction between users and items is an individual's behavior. 
% We now show recent work from the general recommendation system~\cite{} because one can simply treat the sequence as standard features and learn models based on them to fulfill recommendations. 

\emph{User interest model:}
Another area of research that has attracted much attention is user interest modeling, that is, how to model user preferences based on the user’s historical click sequence, which can be applied to click-through rate (CTR) estimation models. 
Deep learning based methods have achieved competitive results in CTR tasks, and recent works typically focus on modeling user latent interest from user historical click sequences~\cite{zhou2018deep,DIEN}. % SIM
DIN~\cite{zhou2018deep} employees the attention mechanism to model target items capturing the differentiated interests of users. 
DIEN~\cite{DIEN} proposes an interest extraction layer to capture the temporal interest from a historical behavior sequence and adopts an interest evolving layer to model the interest evolving process that is relative to the target item. 
Unlike user behavior sequences in E-commerce/advertising scenes, user interactive behaviors in online games are relatively sparse, but indeed there are relatively rich game behavior sequences. 
Such classical user interest CTR models cannot be directly applied to friend application scenarios. 
How to utilize sequences of matchmaking/combat behaviors in game scenarios to recommend strangers is the focus of our research in this paper. 
% \todo{sth missing here}

% \emph{Link prediction:}

% Epasto etc.~\cite{EpastoVLDB2015} developed an community based solution that cluster users into 
% GNNs can be formulated as a message passing framework where node representations are learned by propagating features from local graph neighborhoods via trainable neighbor aggregators.

% \noindent\emph{Social Recommendation:}
% Social recommendation is special type of traditional recommendation that takes advantage of online social relations for enhancing the representation of users~\cite{Tang2013Socialrecommendation}.
% % \todo{more papers needed !}
% \noindent\emph{User similarity:}

% \noindent\emph{User similarity:}
% To get the similarity between users, prior efforts often focuses on the cases where sufficient interactions are available, such as location-based mobile~\cite{Chu2013}, 
% %web information~\cite{ADAMIC2003},
% mutual social influence~\cite{Zheng2017FriendRI}, 
% scholar interaction~\cite{XU2019} and others~\cite{LI2020}.
% The difference in our problem is that we recommend strangers to users without noticeable interactions. Thus, these can not be applied to our task directly.

% \section{User similarity}
% \emph{Link prediction}

% \emph{Network inference}

\begin{table} [t]
\setlength{\abovecaptionskip}{0.3cm}
\centering
 \small
  \footnotesize
\caption{Notation }
\begin{tabular}{llrrrrrrrrr}
%\hline
%\toprule[1pt]
\toprule
\multirow{1}{*}{Symbols} & \multirow{1}{*}{Definitions and Descriptions} & \\          

\midrule

    $p$        &         rating prediction       \\
    $\mathbf{S}_{u_i}$                &  activity sequences  of user i\\ % (x^{(1)}, x^{(2)},..., x^{(t)})      \\
    $\mathbf{H^{(t)}}$                &  user's hidden state representation at first layer at time $t$            \\
    $\mathbf{R^{(t)}_{u_i}}$               &    user's profile feature at time $t$    \\
    $l_{i,j}$ & pairwise link features between $u_i$ and $u_j$\\
    $e^{(t)}$ & user's embedding after feature concatenation at time $t$ \\ 
    $M(\mathbf{e_i,e_j})$ & multi-view similarity between $u_i$ and $u_j$ \\
    % $\Phi(S^{(1)}, S^{(2)},..., S^{(t)})$                & user-pairs' similarity sequences         \\
    $g(\cdot)$                & user-to-user top-level embedding       \\

\bottomrule
\end{tabular}\label{tbl:notation}
% \label{table_2}
% \vspace{-2mm}
\end{table}

% \input{tex/3_prelim.tex}

% \input{tex/4_method.tex}
% \begin{table}[!t]
% \centering
% %\scriptsize
% \begin{tabular}{ |l|c|c|}
%  \hline
% $\mathbf{X}$ &   \\
%  \hline
% $\mathbf{H}$  &    \\
%  \hline
% $\mathbf{C}$  &    \\
%   \hline
% $\mathbf{r}$  &    \\
%   \hline  
% $\mathbf{u}$  &    \\
%   \hline  
%  $\mathbf{Q}$  &    \\
%   \hline  
%   $\mathbf{K}$  &    \\
%   \hline  
% $W$  &   similarity matrix \\
%   \hline  
%  $\odot $ & The Hadamard product  \\
%   \hline
% \end {tabular}
% \caption{ Key notation used throughout the paper \todo{add all major symbols}}
% \label{table:notation}
% \end{table}

\begin{figure*}[t]
    \centering
    \small
    \captionsetup{font={small}}
    \setlength{\abovecaptionskip}{0pt}
    \setlength{\belowcaptionskip}{2pt}
	\includegraphics[scale=0.3]{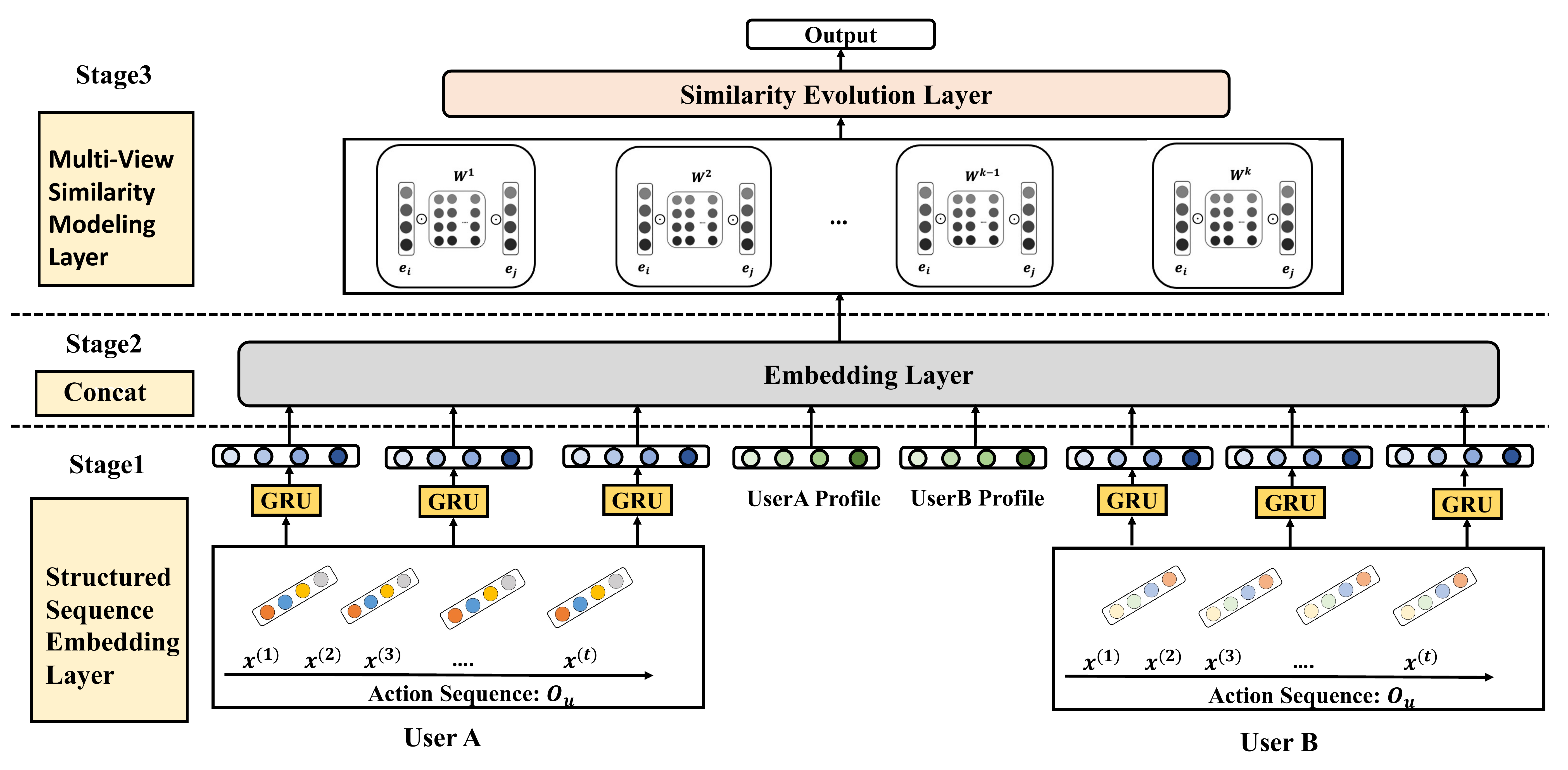}
	\caption{
	%\textcolor{red}{
	%\footnotesize
	An overview of the proposed architecture (DSEN). It consists of three stages. First, in structured sequence Embedding layer, it embeds a series of user action sequences into low-dimensional vectors.  This is then used to concatenate with users' profile vectors to obtain users' embeddings at each time point. At the third stage, we introduce a multi-view similarity model to capture the evolution process of user-user similarity from the obtained users' representations. In this layer, $W_i, \forall i \in {1,2,..k}$ are all learnable parameters, where $k$ is the number of views in similarity vector. Here, $e_i, e_j$ are  the hidden layer embeddings of source and target users, respectively.
We then employ the recurrent structure to understand the similarity between users, which is designed to capture the relevance of similarity from this evolutionary process. 
%}
% 	The overview architecture of the proposed DSEN.  We use classical two-tower structure to construct our deep model, which has three components. First layer is structured sequence embedding layer, which receive a series of user's active sequence and convert the temporal dependency into a hidden layer representation and concatenate it with the user's profile vector to get the user's embedding at each time point. Second layer is multi-view similarity layer, we adopt a group of learnable weight matrices to describe the potential similarity of user-pairs at each historical moment. The last layer is similarity evolution layer which utilize a recurrent-based structure to model the evolution of the similarity between user-pairs.
	}
    \label{DMGAN} 
\end{figure*} ~\label{fig:framework} % \vsa \vsa

% \todo{rui: double check this part:}
% \textcolor{red}{The sequences' embedding is unique but not independent, each  i think it can not be trained in a parallel style  }
% Note that sequences' embedding is independent to each other, therefore, we can do a parallel style training through a distributed system.  This is particularly attractive for large scale datasets, where popular applications often have millions users.
% \todo{next,we need to tell how to do the inference.}

% A notable property of our architecture is that, while our model is trained end-to-end, the item embedding computation is independent for each item. Since most of the computation is done during item embedding, we can pre-compute such embeddings in an offline process. Then, during inference, we fetch relevant item embeddings and only compute the buyer and context embeddings on- the-fly. This is particularly attractive in a production environment, since inference can be done efficiently online without compromising on up-to-date buyer purchase histories.

\section{Deep Neural Friend suggestion }

\subsection{Problem Formulation }
% In this section we first introduce the definition of friend suggestion in our setting as follows,
% % \vspace{0.1cm}

% In a online game platform, an individual user $u_i$ has both long-term \textcolor{red}{personalization} and short-term personalization, denoting as $R_i$ and $S_i$, respectively. The short-term personalization is from the user's instant online activities. For players in a game,
% this contains their actions in a game, such as tactical skills, moving traces and so on. Formally, these can be represented as sequences that collecting from $t$ time slots: $\mathbf{S}_i=  [x^{(1)}, x^{(2)},..., x^{(t)}]$, %providing a refined granularity to $\mathbf{H^{(t)}}$
% capturing the evolution of users' actions over time. 
% Whereas, long-term personalization often includes features that change is slow or constant over a long time period, such as profiles. 
% We integrate these two types of features to describe a user
% while maintaining  time sensitivity.
% In particular, we know little about the social relations between users, and the interactions between users are often sparse in our setting, which is often the case of suggesting strangers to users in games. 
% This setting is rarely considered in the literature, therefore existing solutions are often not applicable to this case. 
% This work aims to address the friend suggestion problem in the just-mentioned setting with user features from both long-term changing and short-term changing perspectives. 

In this section, we first introduce multiple concepts related to friend suggestions in an online game platform. In a large-scale multiplayer online game,  individual user generates a series of their in-game behaviors (recharge, consumption, gaming and social interaction, etc.) are recorded and stored daily, to generate the users' personalization features.
%Users will generate a series of registration data and behavior logs in the game.
We model these data to produce users' profiles to express users' long-term personalization and short-term personalization. The short-term personalization is from the user's instant online activities. 
For players in a game,
this contains their actions in a game, such as tactical skills, moving traces and so on. Formally, these can be represented as sequences that collecting from $t$ time slots: $\mathbf{S}_i=  [x^{(1)}, x^{(2)},..., x^{(t)}]$, %providing a refined granularity to $\mathbf{H^{(t)}}$
capturing the evolution of users' actions over time. 
Whereas, long-term personalization often includes features that change is slow or constant over a long time period, such as profiles. 
In addition to the user's own characteristics, the static interaction relationship between users, such as the number of common friends, the number of games, and the common team, etc., can reflect the close relationship between pairs of users, called pairwise link features.

In particular, we know little about the  social relations between users, and the dynamic interactions between users are often sparse in our setting, which is often the case of suggesting strangers to users in games. 
This setting is rarely considered in the literature, therefore existing solutions are often not applicable to this case. 
This work aims to address the friend suggestion problem in the just-mentioned setting with user features from i) long-term changing, ii) short-term changing perspectives, and iii) pairwise link features. 
%The mathematical symbols used in this paper are given in Table 1.

As mentioned in the last section, users' interaction behaviors play an importance role to enhance  users' experience, increase the game time, and promote consumption. 
More specifically, we extract a pair of interacting users from a friend list directly, or read from the temporary interaction data: sending out and approving friend application. 
In our scenario, we recommend strangers to users as friend lists daily, 
%Since there is a inherent time delay in approving/rejecting friend applications, 
and we employ   the click-through rate of applications as the measurement.

\begin{mdframed}
%[linecolor=red!60!black,backgroundcolor=gray!20,linewidth=2pt,    topline=true,rightline=false, leftline=false]
{\bf Friend suggestion:} 
% Given a user query  and a list of candidates, along with all their individual long-term features  and the short-term features, we are asked to rank the target in the candidate set so as to suggest the proper friends to the user.  
% \par
Given a user $u_i$ and his/her long-term features  $R_{u_{i}}$ and short-term features $S_{u_{i}}$, link features $l_{i,j}$, aiming at learning a function $f(S_{u_{i}}, R_{u_{i}}, l_{i,j})\rightarrow p_{ij}$ for each candidate user $u_j$ where $p_{ij}$ denotes the estimated strength of the potential interaction between $u_i$ and $u_j$. 
This estimated strength represents the likelihood that user $u_i$ tends to interact with user $u_j$ in the future.
Thus, the higher the strength, the more likely user $u_i$ is to make a friend request to user $u_j$. Then, we utilize the estimations to rank the target in the candidate set so as to suggest the proper friends to the user.
% We address this by measuring the similarity between $A$ and targets in $G$ pair-wise as $\left ( A,G_i \right )$.  Our proposed model is to calculate the pairwise similarity as $\mathcal{F}\left ( u_a,u_b \right )$.
\end{mdframed}

\subsection{Methodology}
We now present the details of our proposed framework,
Deep Similarity Evolutionary neural Network (\ourmeth), for friend suggestions, which exploits users' long-term and short-term information. 
The proposed model has the following main components, including the structured sequence embedding layer, the multi-view similarity modeling layer, and the similarity evolution layer.

% We first present the general NCF framework, elaborating how to learn NCF with a probabilistic model that em- phasizes the binary property of implicit data. We then show that MF can be expressed and generalized under NCF. To explore DNNs for collaborative filtering, we then pro- pose an instantiation of NCF, using a multi-layer perceptron (MLP) to learn the user–item interaction function. Lastly, we present a new neural matrix factorization model, which ensembles MF and MLP under the NCF framework; it uni- fies the strengths of linearity of MF and non-linearity of MLP for modelling the user–item latent structures.

\subsubsection{General Framework}
In this paper, we propose a novel friend suggestion framework by ranking people in the candidate set after retrieval through a multi-stage deep structure. 
% %, underlining the time dependency and multiple perspectives of sequence data of online users. 
An overview of our proposed model is shown in Figure ~\ref{fig:framework}.
% In the following we will present the details 
% implement three layers, which are user activity  embedding layer embedding layer, multi-perspective similarity learning layer and sequence based prediction layer. 
The embedding layer takes user's action sequences as input and projects them into a dense vector for each time point, which uses hidden states as output. 
Then the output embedding of each time slot is concatenated with the user's static profile vector, which can comprehensively capture user's long-term and short-term personality. 
Subsequently, the obtained dense features are fed into the next layer to obtain users' historical interests from multi-view viewpoints. 
Finally, we apply an evolutionary similarity layer to output a score to determine the similarity.

% \subsubsection{User activity  embedding layer}
\subsubsection{ Structured sequence embedding layer.}
% \subsubsection{Gated Recurrent Unit embedding}
% \par
% in order to model the input sequences, we apply sequence-aware neural networks to
% The most distinctive characteristic of our setting is to take pair-wise sequences
\eat{
\textcolor{red}{Different from traditional e-commerce recommendations or news recommendations, in a large number of social apps and game apps, the user’s friendship behavior is sparse and random, which is not suitable for modeling user's friend interests via user's interaction sequence with strangers.
In these scenario, the user’s interaction with the app and activity in the app is very rich even if it has less relation with user's friend preference.
Thus, how to utilize these behavior to conduct friend recommendation is vital and difficult.}\todo{this part should not be here. the model section should tell what model looks like, as to the reason/motivation should be at somewhere else.}
}

Individual players constantly output a series of actions in the game, providing rich information to understand the players' state over time. 
The friend suggestion setup needs to understand the recent changes made by the user. 
We therefore start by modeling activity sequences which has directly or indirectly related to the user's friend application preference. 
First, we transform the user's activity sequences $\mathbf{S} (x^{(1)}, x^{(2)}, ..., x^{(t)})$ into a dense representation via an embedding layer, where we take the most up-to-date  $t$ time steps from the user's records and pad the sequence with zeros if one does not have enough data. 
We know that RNN and its variants have the ability to capture inherent sequential dependency, hence we implement this style-based embedding layer to extract features preserving user characteristics. 
% For modeling the continuous behavior of user during the game, such as matches and battles.
We adopt GRU~\cite{chung2014empirical} here because it alleviates the vanishing gradient problem and does not yield more learnable parameters as in RNNs, which can be stated as follows,
% to capture the inherent sequential dependency and obtain the user embedding which can reflect the users' interests. 
% Compared with Recurrent Neural Network (RNN), GRU can alleviates the
% problem of vanishing gradient and does not bring more learnable parameters. 
% More specifically, GRU in our framework can be expressed as:
% \begin{align}
%     \mathbf{u^{(t)}} &= {\rm sigmoid}(g([\mathbf{X^{(t)}};\mathbf{H^{(t-1)}}] + b_u),
%     \notag
%     \\
%     \mathbf{r^{(t)}} &= {\rm sigmoid}(g([\mathbf{X^{(t)}};\mathbf{H^{(t-1)}}] + b_r),
%     \notag
%     \\
%     \mathbf{C^{t}} &= {\rm tanh}(g([\mathbf{X^{(t)}};\mathbf{r^{(t)}\cdot H^{(t-1)}}] + b_c),
%     \notag
%     \\
%     \mathbf{H}^{(t)} &= \mathbf{u}^{(t)} \odot \mathbf{H}^{(t-1)} + (1 - \mathbf{u}^{(t)}) \odot \mathbf{C}^{(t)},
%     \notag
% \end{align}
\begin{equation}
    \begin{cases}
\mathbf{u}^{(\mathbf{t})}= \sigma  \left(g\left(\left[\mathbf{X}^{(\mathbf{t})} ; \mathbf{H}^{(\mathbf{t}-\mathbf{1})}\right]+b_{u}\right)\right.\\
\mathbf{r}^{(\mathbf{t})}=\sigma \left(g\left(\left[\mathbf{X}^{(\mathbf{t})} ; \mathbf{H}^{(\mathbf{t}-\mathbf{1})}\right]+b_{r}\right)\right.\\
\mathbf{C}^{\mathbf{(t)}} =\tanh \left(g\left(\left[\mathbf{X}^{(\mathbf{t})} ; \mathbf{r}^{(\mathbf{t})} \cdot \mathbf{H}^{(\mathbf{t}-\mathbf{1})}\right]+b_{c}\right)\right.\\
\mathbf{H}^{(t)} =\mathbf{u}^{(t)} \odot \mathbf{H}^{(t-1)}+\left(1-\mathbf{u}^{(t)}\right) \odot \mathbf{C}^{(t)}
\end{cases}
\end{equation}
where $\mathbf{X}^{(t)}$ and  $\mathbf{H}^{(t)}$ represent the input and output at time step $t$; 
% $\mathbf{X}^{(t)} \in $ $\mathbb{R}^{d \times 1}$ represents the values of different behavior sequences at time $t$, where $d$ is dimension of input features;
$g(\cdot)$ signifies the fully connected function; $\sigma $ denotes the sigmoid activation function; $ \odot$ is the Hadamard product operation. Besides,  $\mathbf{u}^{(\mathbf{t})}$ and $\mathbf{r}^{(\mathbf{t})}$ are the update gate at $t$ time step and reset gate at $t$ time step, respectively. $\mathbf{C}^{(t)}$ is candidate activation, which is used to receive the information from input state and hidden state.
\par
In this layer, each encoder unit extracts the feature vectors of an input sequence and projects the temporal dependency into a hidden representation. 
Therefore, each user has a unique representation vector at each time slot, which can reflect activity such as skill dynamics in a game. 

After obtaining the embedding for each time slot, we combine it with the long-term features of the players, i.e., profiles. 
The underlying assumption is that one intermediate action of a player may deviate from his/her normal patterns and mislead the similarity estimation process.  Meanwhile, the long-term information alone may not always reflect the player's current preferences. 
It is therefore necessary to combine these two together as an input for computing the similarity at each time step. 
For each user, the hidden state representation is $\mathbf{H^{(t)}}$ and the user's profile features is  $\mathbf{R^{(t)}}$, and the concatenation output is $\mathbf{e^{(t)}}$.

\subsubsection{Multi-view similarity modeling layer}
From the previous step of our model, we obtain a time-aware representation of the players. 
We now introduce a new simple yet effective method for computing the similarity between users. 
% through multiple perspectives.
% Given that we intend to quantize the similarity between users, we introduce a new simple but effective method to compute this through multiple perspectives.
% We now present a generalization solution for measuring the similarity between the embedding features.  
Often, recommendation systems use dot product to get the similarity between input embedding pairs $(\mathbf{e_i}, \mathbf{e_j})$, denoting $ \mathbf{e_i^T} \mathbf{e_j}$. 
With simply linear algebra knowledge, we know that this score function can be formulated as a bilinear form of $\mathbf{e_i^T I e_j}$, where $\mathbf{I}$ is an identity matrix. However, this undermines the importance of the similarity function since the similarity measure and embedding should be mutually enforced to achieve recommendation success. 
This abandons the discriminative power of the similarity function by using the identity matrix, making the prediction models fully dependent on the embedding. 
To overcome this limit, we choose to learn  a parametric similarity subspace, $\mathbf{W}\in\mathbb{R}^{m \times m}$, to better fit the data, resulting in a general similarity measurement as $M(\mathbf{e_i,e_j}) =\mathbf{e_i^T W e_j}$.  
Also, learning a single similarity matrix may not be able to fully capture the relations between input feature-pair embeddings. 
Inspired by ensemble learning theory, we aim to learn a group similarity function to determine similarity collectively. More specifically, we can formulate ensemble similarity learning as 
% $ M(\mathbf{e_i,e_j}) =\sum_k\mathbf{e_i^TW^ke_j}$,
% \vsa 
\begin{equation}~\label{VIEW_FUNCTION}
 M(\mathbf{e_i,e_j}) =\sum_k\mathbf{e_i^TW^ke_j}
 %= \sum_k W^k\otimes \left ( e_i^Te_j \right )=\sum_k\sum_{pq}W_{pq}^ke_{ip}e_{jq}
\end{equation} 
where $\bf W^k$ denotes the $k$th similarity matrix. To implement this, we can further rewrite this bilinear into a linear format like the following:
\begin{equation}
\begin{aligned}
 &M(\mathbf{e_i,e_j}) =\sum_k\mathbf{e_i^TW^ke_j} = \sum_k \mathbf{W^k}\otimes \left (\mathbf{e_i^Te_j} \right )\\
 &=\sum_k\sum_{pq}\mathbf{W_{pq}^ke_{ip}e_{jq}}
 \end{aligned}
\end{equation}
By vectorization, we have $ \mathbf{z} = \left ( \mathbf{e_{i1}}*\mathbf{e_{j1}},...,\mathbf{e_{im}}*\mathbf{e_{jm}} \right )^T$ and $\mathbf{v^k} = \left ( \mathbf{W_{11}^k},\mathbf{W_{21}^k}...,\mathbf{W_{mm}^k} \right )^T$, and reformulate the similarity measurement as follows,
\begin{equation}
     M(\mathbf{e_i,e_j}) =\sum_k\mathbf{e_i^TW^ke_j} = \left [\mathbf{v^1},\mathbf{v^2},...,\mathbf{v^k} \right ]^T\mathbf{z} % =S^Tz
\end{equation}
This can be implemented as a standard Multi-Layer Perceptron (MLP) network having
\begin{equation}
    \begin{aligned}
    % &\mathbf{z}_{1} = [\mathbf{x},\mathbf{y}]\\
    % &....\\
\mathbf{g} &=u\left(\mathbf{V}^{T}  {\mathbf{z}}+{b}\right)  \\
% &\hat{y} &=\sigma\left(\mathbf{h}^{T} \phi_{L}\left({z}_{L-1}\right)\right)
\end{aligned}
\end{equation}
where $b$ is the bias term and $\mathbf{V} = \left [\mathbf{v^1},\mathbf{v^2},...,\mathbf{v^k} \right ]$. Here, $u()$ is activation function, and we use ReLU to promote sparsity in outputs.

% \todo{we can decide whether or not add this bias term of mlp, may be we should?}
% To fully exploit the information in features, we also consider a naive combination of features from both users, denoted as $\mathbf{b}_{L} =V_L^T[e_i; e_j]$ 

% we propose a sequential loss function for this purpose, which is written as 
% \begin{equation}
%     \begin{aligned}
%         l(\mathbf{u}_i, y_{i})= \lambda p\left(y_{i} \mid \mathbf{u}_{i}\right)+(1-\lambda)  q\left(z_{i} \mid z_{i-k}, \cdots, z_{i-1}\right)
%     \end{aligned}
% \end{equation}

% \subsubsection{Similarity evolution sequence}

% \subsection{Sequence based prediction }

\subsubsection{Similarity evolution modeling layer.}
% \textcolor{red}{After we obtain the similarity sequence via multi-view similarity modeling layer, we aim to learn a user-pairs similarity pattern from this sequences.}
So far we have obtained the similarity embedding at each time point for a given user-user pairs. 
The goal of this layer is to produce user-user pair similarity based on the output from the multi-view similarity modeling layer. 
As stated before, the similarity between a user-user pair will evolve over time. 
For example, online game players will change their strategy of adding strangers according to their own level changes, and naturally the similarity between users will evolve over time.  
The historical similarities will have different relevance effects on the current time slot, therefore the problem now translates into capturing this underlying relevance. 
While memory mechanism has been shown to be an effective way of capturing relevant information, we therefore employ it to learn the underlying similarity evolution automatically. 
Specifically, we implement our similarity evolution modeling layer by leveraging LSTM~\cite{2014Learning}, in which the similarity sequence obtained from the multi-view similarity modeling layer is used as input to the LSTM. 
Each LSTM unit receives a similarity vector and transforms the user-pair temporal dependency similarity into a hidden representation. 
In this way, our similarity evolution layer can capture the inherent relations between historical behavior sequences. 

\subsubsection{Prediction layer}
In the similarity evolution layer, we obtain a representation vector $\mathbf{g}$ that characterizes the evolution of the users similarity. 
We then adopt a single-layer feed-forward neural network to obtain the similarity scores between user-pairs at the current time step: 
$p =\sigma \left(\mathbf{W}^{T}  {\mathbf{g}}+{b}\right)$,
% \begin{equation}\label{predict}
%   p &=\sigma \left(\mathbf{W}^{T}  {\mathbf{g}}+{b}\right)  \\
% \end{equation}
where $p$ is the rating prediction, $\mathbf{W}$ is the learnable parameters, $b$ is the bias term, $\sigma()$ is the activation function and we choose Sigmoid in this work.

\subsubsection{Loss function.}
To predict whether a user-pair will establish a friend-request connection, we model it as a binary classification task, and use sigmoid function as the output unit. 
In this work, we use the cross-entropy loss to train our model:
\begin{equation}\label{loss_func}
    \mathcal{L}(\theta)=-\frac{1}{N}\sum_{(x,y)\in \mathcal{D}} (y\log p(\theta) + (1-y)\log(1-p(\theta))
\end{equation}
where $\mathcal{D}$ denotes all the training samples and $y \in \{0,1\}$ is the ground truth label whether user A has sent a friend request to user B,  $p(\theta)$ 
is the output of our framework which represents the prediction probability.

\subsubsection{Sampling strategy}
In our scenario, we recommend each user a list of candidate, also called the exposed list.
The positive samples are defined as these pairs that users click send the friend request to others.
All these pairs have no request action are viewed as negative samples,
and we sample from these due to reduce the computational cost and balance the positive samples.
In general, we keep the ratio between  positive instances and negative instances as $1:4$.
Although the friendship behavior in the game is sparse, 
we can 
due to the large number of active users in our game, a certain amount of friend request logs are still generated daily.
Thus, in our industrial settings, such samples are commonly extracted by routine batch jobs and populated in an upstream database at regular time intervals, to facilitate efficient model training.

\subsubsection{Candidate retrieval.}
% \textcolor{red}{
In our scenario, friend suggestion includes two key sections: the candidate retrieval and the friend ranking, which is coarse to refine process. 
Candidate retrieval aims to generate a list of top-N (e.g., N = 1000) potential friends out of dozens millions users.
Then friend ranking is to implement fine-grained re-ranking from the generated candidates, delivering top-n (e.g., n = 100) users to end users. 
% The candidate retrieval methods must satisfy two requirements: selected candidates should have high potentiality to establish connection from $u_i$ and should
% be suitable for calculation method to ensure acceptable computation complexity.
We follow popular strategy in this domain to do the candidate retrieval, which is multiple way retrieval.
% \todo{lin:add our recall methods}
% \textcolor{red}{
Furthermore, in our scenario, we only suggest $100$ strangers for each user per day, so there is no need to support real-time retrieve and online inference.

% Considering the matter of positions of each word, we utilize position encoding as in~\cite{2017Attention} to represent the sequence relationship.

% \textbf{Output or classifier:}
% \todo{we need this section, right? }
% \textcolor{red}{In this section, it is only a linear projection layer with a sigmoid activation function, so we just mention it is enough?}

\section{Experiment }

\begin{table}
\centering
\begin{tabular}{|l|l|l|l|l|l|} 
\hline
Dataset  & Section-1 & Section-2  & Section-3 & Section-4 \\ 
\hline
% \# test users                   &  5.5 M   & 2.5M & 5.7M & 5.7M  \\
\# train user-user pairs                &  4M     & 11.4M & 4.1M & 14.5M  \\
\# train friendship pairs             & 85k   & 2.7M & 80k & 4.3M  \\
\# test user-user pairs  &  16.1M  & 4.6M & 17.3M  & 14.6M \\
\# sequences features             &  55  &  55 & 55 & 55\\
\# profiles    features           &  68  &  68 & 68 & 68 \\
\hline
\end{tabular}
 \caption{  The  statistics of Dataset I. }~\label{tbl:dataset}
\end{table} % \vsa 

%%%%%%%%%%
%%%%%%%%%%
%%%%%%%%%%

\begin{table}
\centering
\begin{tabular}{|l|l|l|l|l|l|} 
\hline
Dataset  & Section-5 & Section-6  \\ 
\hline
% \# test users                   &  5.5 M   & 2.5M & 5.7M & 5.7M  \\
\# train user-user pairs                & 5.0M    &  9.1M  \\
\# train friendship pairs          & 1.1M   &  1.9M  \\
\# test user-user pairs               &  4.9M  & 7.1M  \\
\# sequences features             &  55  &  55 \\
\# profiles    features           &  68  &  68 \\
\hline
\end{tabular}
 \caption{ The  statistics of Dataset II. %The statistics of datasets having stable social graphs. 
 }~\label{tbl:dataset_gnn}
\end{table}%\vsa 

\begin{table} [t]
\setlength{\abovecaptionskip}{0.3cm}
\centering
 \small
\caption{In game player features description}
\begin{tabular}{llrrrr}
%\hline
%\toprule[1pt]
\toprule
\multirow{1}{*}{Feature type} & \multirow{1}{*}{Description} & \multirow{1}{*}{Number}  & \\          

\midrule

                    &    Registration information      &  13        \\
Profile
(long-term)
                    &    Game mode    &     14     \\
                      &     Activation summary   &    28      \\
                    &     Consumption summary   &    13      \\
                       
\midrule

                    & Score and award  count  &  10 \\
 Behavior
 (short-term)
                    &   Equipment count         &   4        \\
                    &   Tactical skills         &   37        \\
                    &   Team statistics         &    4       \\

\bottomrule
\end{tabular}\label{tbl:user_feature}
% \label{table_2}
\vspace{-2mm}
\end{table}

%+++++++++++++
%+++++++++++++
%+++++++++++++
%+++++++++++++

\begin{table*} [t]
\setlength{\abovecaptionskip}{0.3cm}
\centering
 \small
\caption{Friend suggestion performance comparison between ours and the competing techniques on four large datasets}
\begin{tabular}{llrrrrrrrrr}
%\hline
%\toprule[1pt]
\toprule
\multirow{2}{*}{Dataset} & \multirow{2}{*}{models} & \multicolumn{2}{c}{K=10}         & \multicolumn{2}{c}{K=20}       & \multicolumn{2}{c}{K=50}    & \multicolumn{2}{c}{K=100}                       \\ \cline{3-10}
                         &                         & HIT@K           & NDCG@K                   & HIT@K           & NDCG@K                 & HIT@K           & NDCG@K         & HIT@K           & NDCG@K              \\ %\hline
\midrule
                        % & HA   & 4.16 & 7.80 &13.0\% & 4.16 & 7.80 &13.0\% & 4.16 & 7.80 &13.0\%
                        % \\
                         & MLP   &  0.0688    & 0.0334     &   0.1063   &  0.0429     &       0.1877 & 0.0589       &    0.2627   &  0.0711            \\
                         
                         & GRU                 &  0.1547    &  0.0987    &   0.2039  &  0.1111     &       0.2733 &  0.1249     &  0.3252     &   0.1333     \\
                   
                         & self-Attention    &   0.1662     &   0.1050   &   0.2189  &  0.1183     &  0.2897    &  0.1324     &    0.3459   &  0.1415     \\
   Section-1  
                       & \textbf{DSEN}          & \textbf{  0.1680}    &   \textbf{0.1071}  &  \textbf{0.2222}   &    \textbf{0.1207}   &       \textbf{0.2986} &    \textbf{0.1359}   &   \textbf{0.3532}    &  \textbf{0.1448}      \\

\midrule
                        % & HA & 2.88 & 5.59 & 6.8\% & 2.88 & 5.59 & 6.8\% & 2.88 & 5.59 & 6.8\%
                        % \\
                         & MLP                   &   0.1780       &        0.1059         &   0.2358       &   0.1204              &         0.3149 &     0.1362 & 0.3773 & 0.1463        \\
                         & GRU                 &     0.1784     & 0.1109                 &      0.2328     &    0.1246               &       0.3142  &       0.1407 & 0.3728 & 0.1502           \\
                         & self-Attention         &    0.1881     &    0.1130              &  0.2464     &    0.1278             & 0.3312         &          0.1446 & 0.3888 & 0.1540     \\
   Section-2  
                       & \textbf{DSEN}              &   \textbf{0.1914 }      &     \textbf{0.1167   }    &     \textbf{0.2499 }      &   \textbf{0.1315 }                &  \textbf{0.3421 }     &      \textbf{0.1498}  & \textbf{0.4050}  & \textbf{0.1600 }            \\
                       
\midrule
                        % & HA & 2.88 & 5.59 & 6.8\% & 2.88 & 5.59 & 6.8\% & 2.88 & 5.59 & 6.8\%
                        % \\
                         & MLP   &     0.1796    &   0.1108             & 0.2334        &   0.1244              & 0.3062  & 0.1389   & 0.3615 &  0.1479       \\
                         & GRU                 &     0.1840     & 0.1173                 &      0. 2330    &    0.1296              &       0.3069  &       0.1444 & 0.3587 & 0.1528           \\
                         & self-Attention         &    0. 1787    &    0.1119              &  0.2286     &    0.1244             & 0.3005         &          0.1387 & 0.3538 & 0.1474     \\
   Section-3  
                       & \textbf{DSEN}              &   \textbf{0.1846 }      &     \textbf{0.1159}    &     \textbf{0.2363 }      &   \textbf{0.1289 }                &  \textbf{0.3117 }     &      \textbf{0.1439}  & \textbf{0.3679}  & \textbf{0.1531}            \\
\midrule
                        % & HA & 2.88 & 5.59 & 6.8\% & 2.88 & 5.59 & 6.8\% & 2.88 & 5.59 & 6.8\%
                        % \\
                         & MLP   &  0.1296      & 0.0812            & 0.1634       &  0.0898               & 0.2134  & 0.0997   &0.2435  &0.1046      \\
                         & GRU                 &     0.1305     & 0.0819      &      0.1656     &    0.0907              &       0.2132  &       0.1002 & 0.2455 & 0.1054           \\
                         & self-Attention         &    0.1291     &    0. 0873             &  0.1644     &    0.0.0929             & 0.2144         &          0.1008 & 0.2461 & 0.1049     \\
   Section-4  
                       & \textbf{DSEN}              &   \textbf{0. 1320}      &     \textbf{0.0.0829}    &     \textbf{0.1698 }      &   \textbf{0.0924 }                &  \textbf{0.2191 }     &      \textbf{0.1022}  & \textbf{0.2527}  & \textbf{0.1077}            \\         
                
\bottomrule
\end{tabular}\label{tbl:COMPARSION}
% \label{table_2}
%\vspace{-2mm}
\end{table*}

\begin{table} [t]
\setlength{\abovecaptionskip}{0.3cm}
\centering
 \small
 %\footnotesize
\caption{
%\footnotesize
The performance comparison between DSEN and GNN.}
\begin{tabular}{llrrrrrrrrr}
%\hline
%\toprule[1pt]
\toprule
\multirow{2}{*}{Dataset} & \multirow{2}{*}{models} & \multicolumn{2}{c}{K=10}     & \multicolumn{2}{c}{K=100}                       \\ \cline{3-6}
                         &                         & HIT@K           & NDCG@K             & HIT@K           & NDCG@K              \\ %\hline
\midrule
                   
                         & GraFRank    &   0.0035     &   0.0182   &     0.1666   &  0.0432     \\
                        %  & MLP    &   0. 1457    &   0.0882   &   0.1876  &  0.0986     &  0.2448   &  0.1101     &    0.2988   &  0.1188     \\
                        %  & GRU    &   0.1429     &   0.0884   &   0.1827  &  0.0985    &  0.2440   &  0.1106     &    0.2977   &  0.1193     \\
                        %  & seld-Attention   &   0. 13    &   0.   &   0.  &  0.    &  0.   &  0.     &    0.   &  0.     \\
  Section-5  % 0926
                      & \textbf{DSEN}          & \textbf{ 0.1568}    &   \textbf{0.0954}  &   \textbf{0.3269}    &  \textbf{0.1296}      \\

\midrule

                         & GraFRank  &    0.051     &    0.024     & 0.2328 & 0.0595     \\
                        %   & MLP    &   0.     &   0.   &   0.  &  0.     &  0.   &  0.     &    0.   &  0.     \\
                        %  & GRU    &   0.  1916   &   0.1203   &   0.2539  &  0.1360     &  0.3342   &  0. 1520    &    0.3980   &  0.1624    \\
  Section-6  %0914
                      & \textbf{DSEN}              &   \textbf{0.1976 }      &     \textbf{0. 1225  }  & \textbf{0.4095}  & \textbf{0.1657 }            \\

\bottomrule
\end{tabular}\label{tbl:COMPARSION_gnn}
% \label{table_2}
%\vspace{-2mm}
\end{table}

%%%%%%%%%%%%%
%%%%%%%%%%%%%
%%%%%%%%%%%%%
%%%%%%%%%%%%%
%%%%%%%%%%%%%

\begin{table} [t]
\setlength{\abovecaptionskip}{0.3cm}
\centering
 \small
 \footnotesize
\caption{
%\footnotesize
We perform ablation analysis on two datasets. Attention-based model has inferior performance than the LSTM-based one. }
\begin{tabular}{llrrrrrrr}
%\hline
%\toprule[1pt]
\toprule
\multirow{2}{*}{Dataset} & \multirow{2}{*}{models} & \multicolumn{2}{c}{K=10}         &  & \multicolumn{2}{c}{K=100}                       \\ \cline{3-6}
                         &                         & HIT@K           & NDCG@K                     & HIT@K           & NDCG@K              \\ %\hline

\midrule

                         & DSEN-ATT         &0.1767    &0.0712    & 0.3688 &0.1062     \\
  Section-2  
                      & \textbf{DSEN}              & \textbf{0.1914 }      &     \textbf{0.1167   }     & \textbf{0.4050}  & \textbf{0.1600 }            \\
                       
\midrule

                         & DSEN-ATT          &0.1633     &0.0675             & 0.3370  & 0.0991     \\
  Section-3  
                      & \textbf{DSEN}              &\textbf{0.1846 }      &     \textbf{0.1159}    & \textbf{0.3679}  & \textbf{0.1531}            \\

\bottomrule
\end{tabular}\label{tbl:COMPARSION_ablation}
% \label{table_2}
\vspace{-2mm}
\end{table}

\subsection{Experimental settings}

\subsubsection{Datasets}
We evaluate \ourmeth on  several large-scale game datasets, shown as Table.~\ref{tbl:dataset}. 
Our data is all collected from a large-scale online  game from a major online game company in the world that contains about six millions of players. 
And we divide the data into different sections (Section1-6) based on logging time and user area.
We employ in-game user profiles and players' actions as long-term and short-term features, respectively. 
More specifically, we have $68$ features from profiles and $55$ features from action features, as shown in Table.~\ref{tbl:user_feature}, where all action features are collected daily from the past $15$ days. 
All features are pre-processed with zero-mean and unit-variance normalization. 
%\textcolor{red}{The details of features description are as follows:}
% \par
% \textbf{Registration information:}
% \par
% \textbf{Game mode:}
% \par
% \textbf{Activation summary:}
% \par
% \textbf{Consumption summary:}
% \par
% \textbf{Score and award count:}
% \par
% \textbf{Equipment count:}
% \par
% \textbf{Tactical skills:}
% \par
% \textbf{Team statistics:}
% \par
% \textbf{pairwise features:}
% \par
% \todo{lin:add more datails}
% Positive samples are pairs that one user has applied to the other. 
% Negative samples are randomly chosen from exposed user-user pairs without requests. 
% For each positive instance, we sampled about four negative instances. 
We used 80\% as training set for learning parameters and 20\% as validation set for hyper-parameter tuning for each training dataset. 
Test dataset is the one collected from the day next to date in training dataset,  ensuring no label exposure.

% For comparison purpose, we have two types of datasets in two tables. The datasets in  Table.~\ref{tbl:dataset_gnn} contain stable social networks that have less cold start users than thoese in Table.~\ref{tbl:dataset}. Therefore, we can implement experiments for graph-based models.
% \textcolor{red}{Two types of datasets are used to compare them in two tables and both types of data are collected from a same game.}
Four types of datasets are used for the comparison, where all these are collected from the same game but different sections.
The datasets in Table.~\ref{tbl:dataset_gnn} can generate stable social networks, which have fewer cold start users than those in Table.~\ref{tbl:dataset}. 
We can therefore implement experiments for graph-based models using datasets in Table.~\ref{tbl:dataset_gnn}.

\begin{figure*}[!t]
    \begin{minipage}[b]{0.3\linewidth}
     \captionsetup{font={small}}
		\centering
		\includegraphics[scale=0.22]{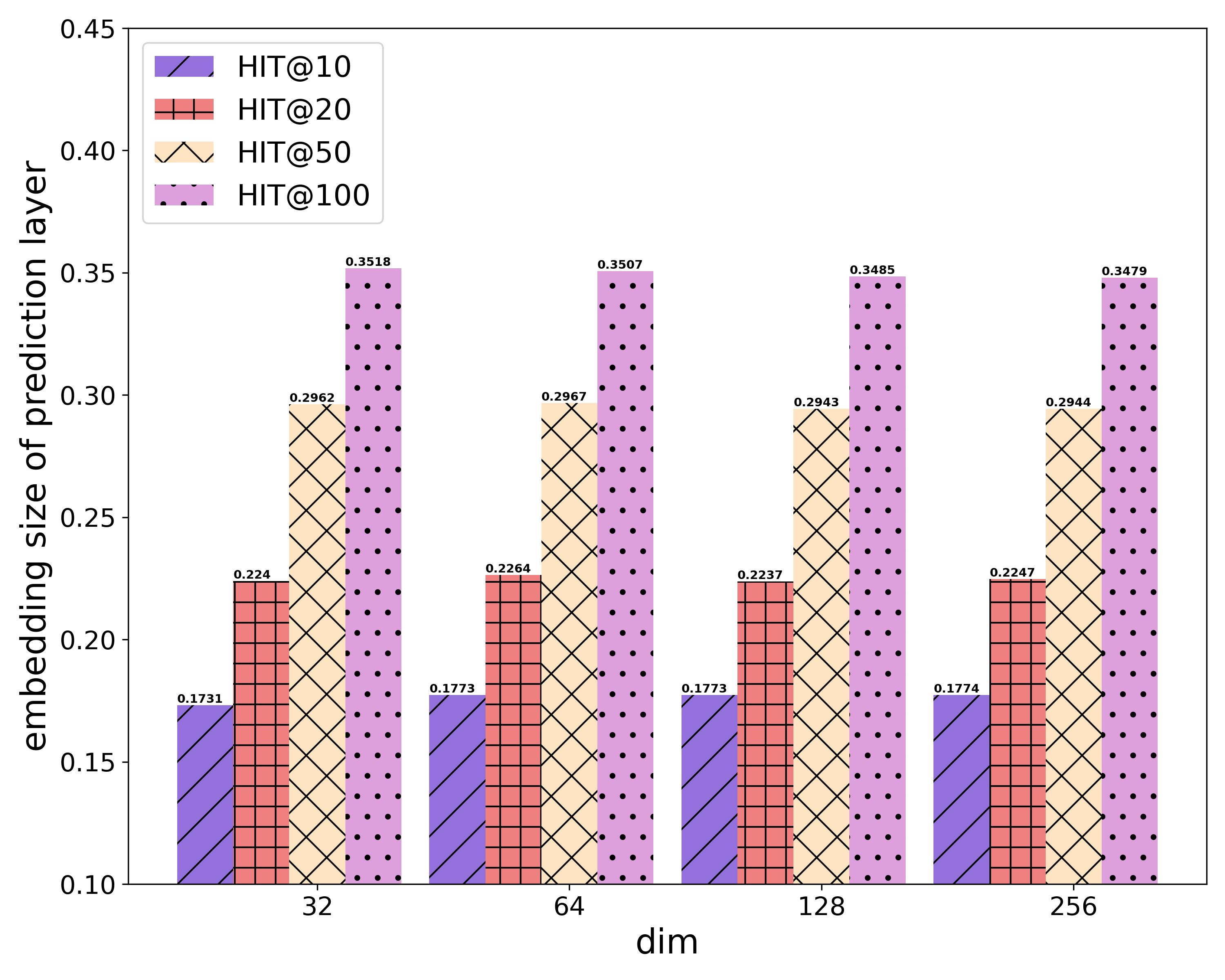}
% 		\centerline{\small{}}
	\end{minipage}
    \vspace{.003in}
    \hspace{.08in}
	\begin{minipage}[b]{0.3\linewidth}
        \captionsetup{font={small}}

		\centering
		\includegraphics[scale=0.22]{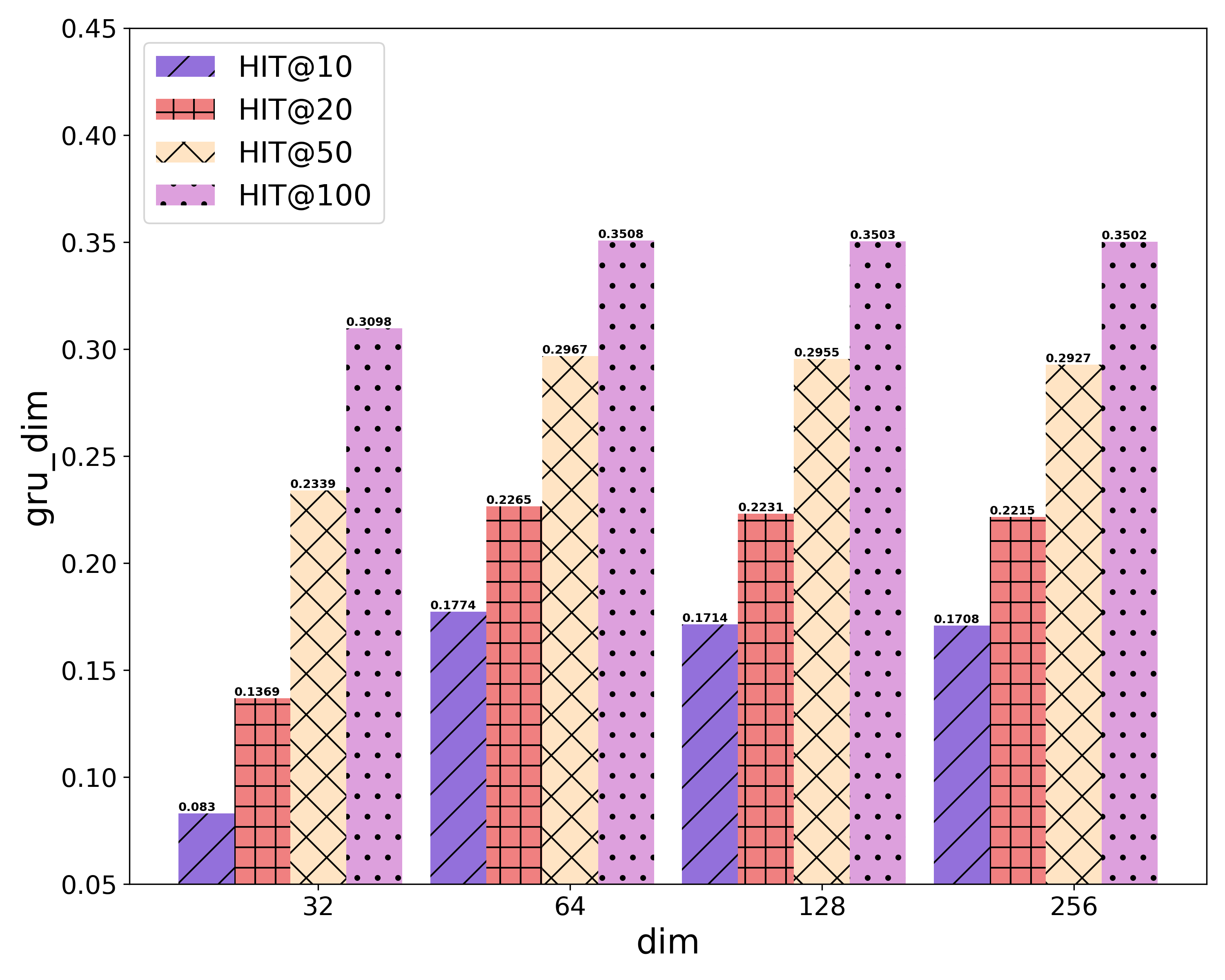}
% 		\centerline{\small{(a) HIT@K}}
	\end{minipage}
    \vspace{.003in}
    \hspace{.08in}
	\begin{minipage}[b]{0.3\linewidth}
        \captionsetup{font={small}}

		\centering
		\includegraphics[scale=0.22]{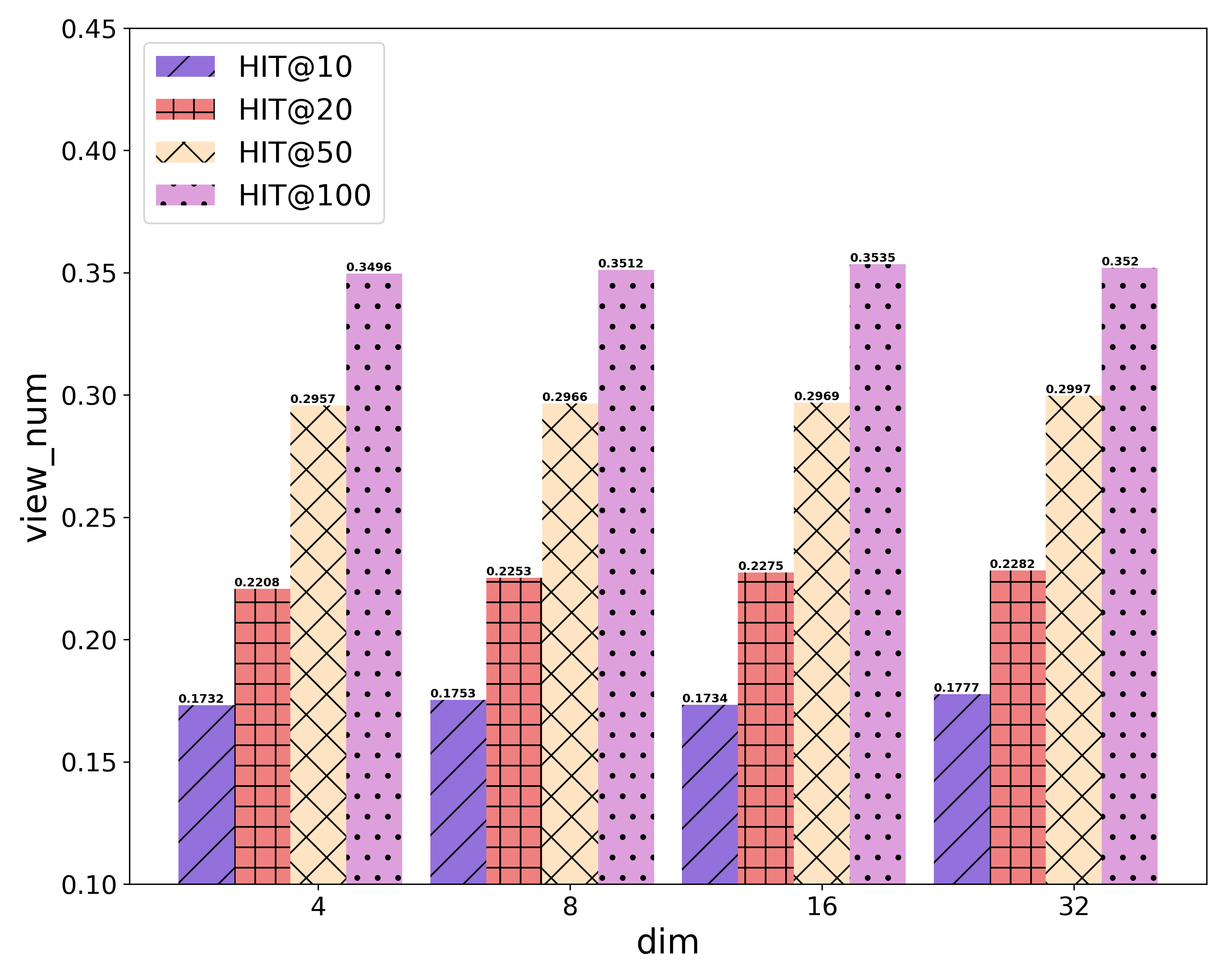}
% 		\centerline{\small{}}
	\end{minipage}
	\vspace{.003in}
    \hspace{.08in}
	
	\caption{
	% \footnotesize
	Configuration Analysis of HIT@K, including embedding size of the prediction layers, GRU dimension and the number of views in similarity.}
	\label{fig_hit}
% 	\vspace{-2mm}
\end{figure*}

\begin{figure*}[!t]
    \begin{minipage}[b]{0.3\linewidth}
     \captionsetup{font={small}}
		\centering
		\includegraphics[scale=0.22]{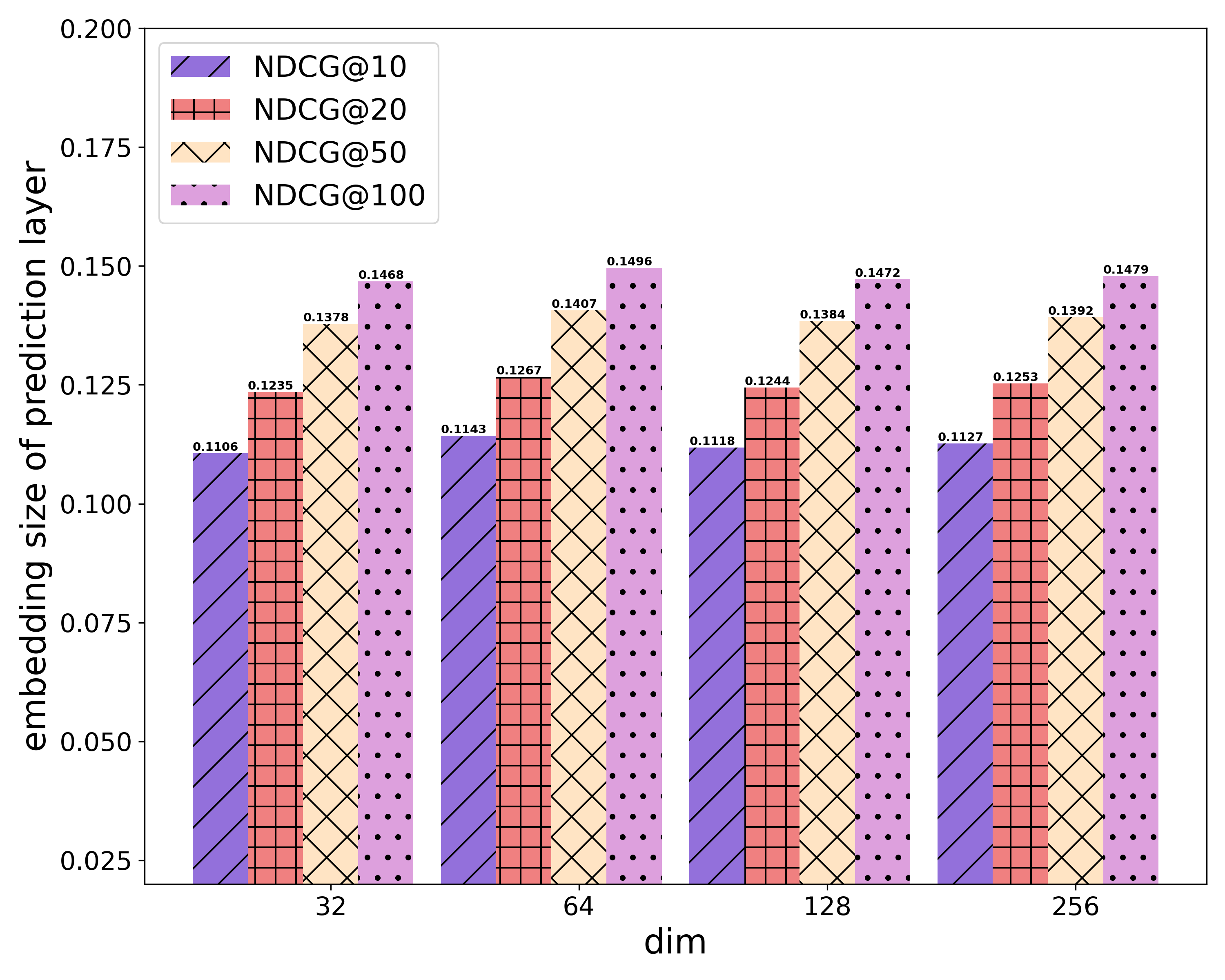}
% 		\centerline{\small{}}
	\end{minipage}
    \vspace{.003in}
    \hspace{.08in}
	\begin{minipage}[b]{0.3\linewidth}
        \captionsetup{font={small}}

		\centering
		\includegraphics[scale=0.22]{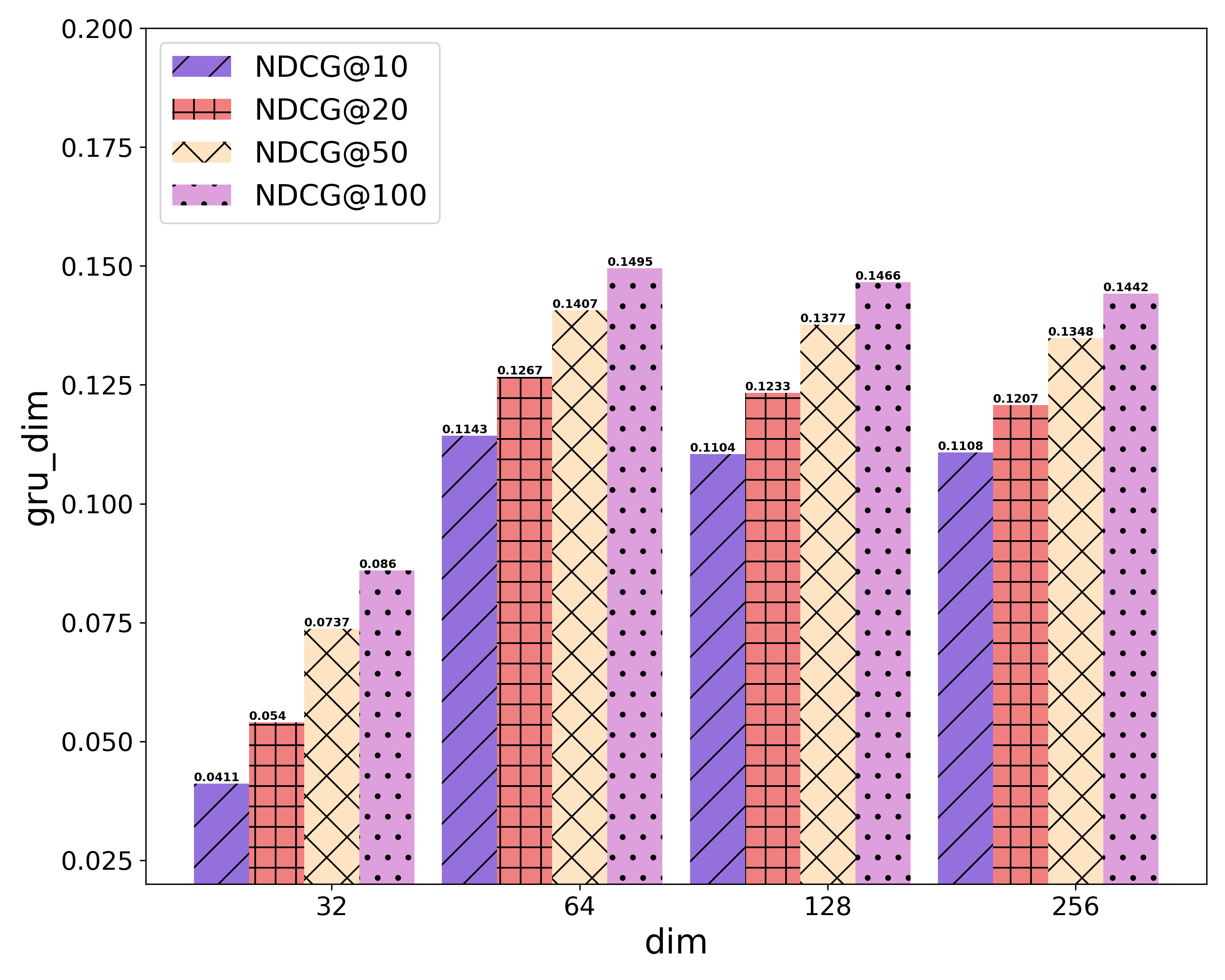}
% 		\centerline{\small{(a) HIT@K}}
	\end{minipage}
    \vspace{.003in}
    \hspace{.08in}
	\begin{minipage}[b]{0.3\linewidth}
        \captionsetup{font={small}}

		\centering
		\includegraphics[scale=0.22]{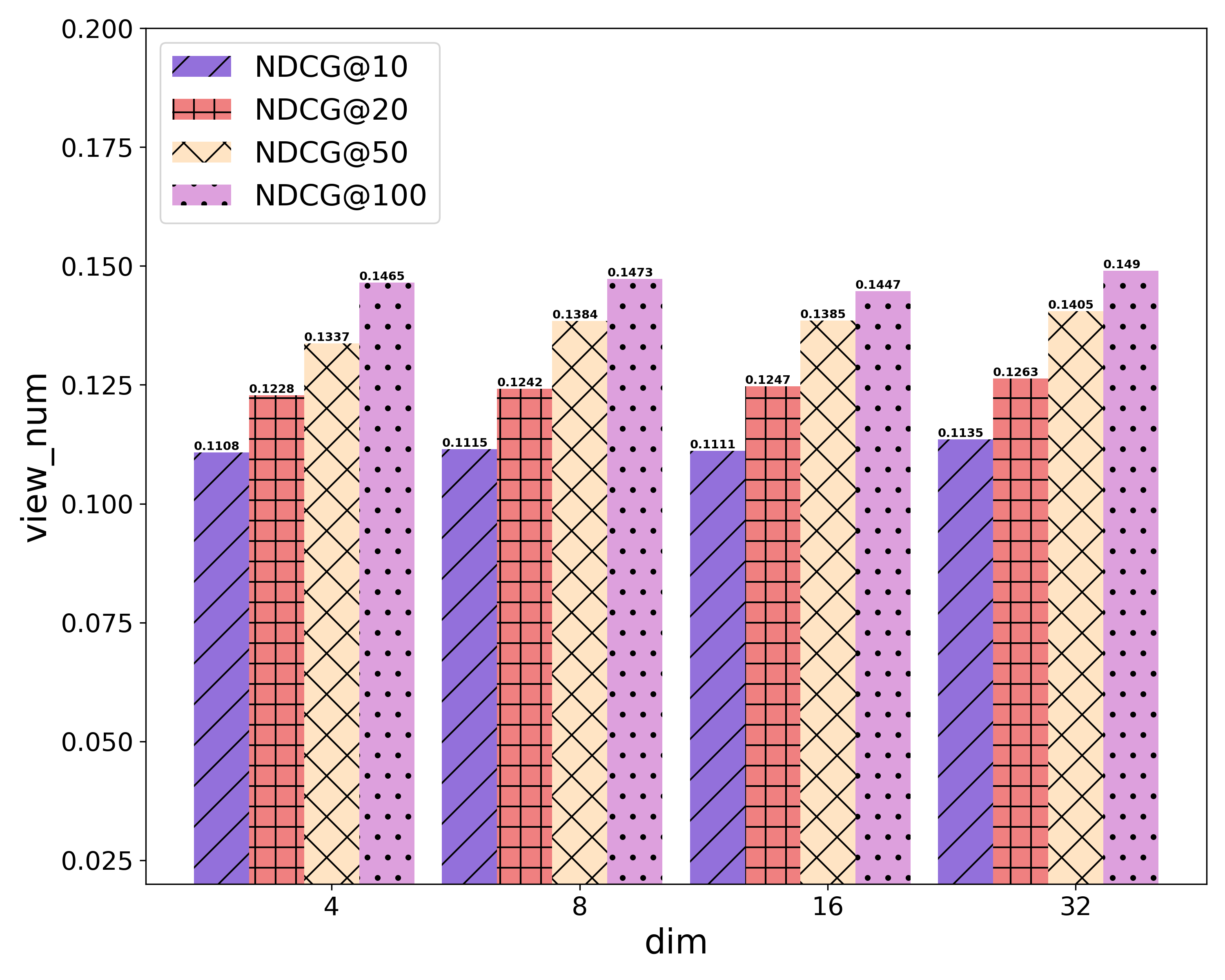}
% 		\centerline{\small{}}
	\end{minipage}
	\vspace{.003in}
    \hspace{.08in}
	
	\caption{
	%\footnotesize
	Configuration Analysis of NDCG@K, including embedding size of the prediction layers, GRU dimension and the number of views in similarity.}
	\label{fig_ndcg}
% 	\vspace{-2mm}
\end{figure*}

\subsubsection{Baselines}

We compared our proposed method to four baselines which are all deep neural network based methods for performance evaluation. 
Since our proposed methods aim to address friend suggestion in online game settings, we do not have any other models can implement this case directly, therefore we compare with models formulating from popular deep neural models such as MLP and GRU.
% \todo{reason why choose this baseline}
% Moreover, our model is a typical two-tower structure, used in the \textit{\bf rough ranking} stage of the entire friend suggestion chain, so we do not compare with the traditional CTR model that is for \textit{\bf refine ranking.}

% \todo{lin:add random rec, ppr?}

{
\setlength{\parskip}{0.5em}
\par
$\bullet$ \textbf{GRU}: 
On the source user tower side and the target user tower side, we adopt Gated Recurrent Unit to extract the short-term interest of the user's active sequence, and the top layer outputs the rating score through a single-layer fully connected network.
~\cite{2014Learning}. 
\noindent
\hangafter=1
\setlength{\hangindent}{2em}
\par
\par
$\bullet$ \textbf{MLP}: Multi-Layer Perceptron with fully-connected
layers and ReLU activations to learn user representations~\cite{2016Deep}. This shares the same ranking loss with the proposed method \ourmeth.
\noindent
\hangafter=1
\setlength{\hangindent}{2em}
\par
% $\bullet$ \textbf{self-attention }: 

\par
$\bullet$ \textbf{self-Attention}: 
% Using self-attention to capture the context dependence
Using a self-attention module to capture the hidden temporal dependency in the user's active sequence, then calculating the current prediction score based on the similarity evolutionary process learned the user-user pairs embedding~\cite{2017Attention}. 
% between sequences simultaneously~\cite{2017Attention}
\noindent
\hangafter=1
\setlength{\hangindent}{2em}
\par

\par
$\bullet$ \textbf{GraFRank}: Learning user representation from social networking, integrating with  multi-modal user feature, for friend suggestion. 
~\cite{SankarWWW2021}.
\noindent
\hangafter=1
\setlength{\hangindent}{2em}
\par

}

Note that GraFRank explicitly depends on the social graph while the features introduced in this paper are independent of the structure of the network.
A direct comparison between our proposed model and GraFRank is therefore unfair. 
We report GraFRank results on a few datasets using the social network.
But such comparisons merely suggest that building a friend suggestion model based on individual independent features is a promising direction as a surrogate of social network design, especially when social network information is insufficient.

% we thus implement a comparison with a state-of-the-art method, called {\bf GraFRank}~\cite{SankarWWW2021}, using GNN. 
% We note that GraFRank has employed different features including social network structure and feature modalities, while \ourmeth uses partial information. To be fair, we only conduct the comparison on one dataset as a reference. 

\subsubsection{Evaluation Metrics.}
To evaluate friend recommendation, we use two popular metrics: Hit-Rate (HR@K), Normalized Discounted Cumulative Gain (NDCG@K), where $K$ denotes the top-chosen number. 
% The definition of NDCG@K is $NDCG@ K=\frac{N D C G_{u} @ K}{|u|}$, where $D C G @ K=\sum_{i}^{K} \frac{r(i)}{\log _{2}(i+1)}$.
Large values of Hit-Rate and NDCG indicate better predictive performance. 
Compared with the HR@K, NDCG@K is more sensitive to the order and position of the recommended list, which can reflect whether the relevance ranking conforms to the target user's real preference priority.
It is worth noting that the number of users is often very large for online applications, \textbf{small improvements to these metrics can have significant impact on overall performance}, thus improving online user experience.

\subsubsection{Parameter Settings.}
We implement our model based on Tensorflow2~\cite{AbaBar16Tensorflow}.
%\todo{lin:training environment}
The whole model is learnt by optimizing the log loss of Eq~\ref{loss_func}. 
For the embedding size $d$, we tested the value in the set of $[ 8, 16, 32, 64, 128, 256 ]$. We search the batch size and learning rate in $[ 1024, 2048, 4096, 8192, 16384]$ and $[ 0.001,0.005, 0.01, 0.05, 0.1 ]$, respectively. 
An early stopping strategy was performed, where we stopped training if the training AUC on the validation set has no increase for two consecutive epochs of training. 
For all learnable parameters in the network, we use random initialization with variance scaling.
% In all neural network methods, we randomly initialized the model parameters with a Gaussian distribution, where the mean and standard deviation are $0$ and $1$ , respectively. 
Moreover, we adopt Adam optimizer to train all learnable parameters.
For the baseline algorithms, the parameters were initialized as in the corresponding papers, and were then carefully tuned to obtain the optimal performance. We also present the values of the key parameters used for the final results at Table~\ref{tbl:param_model}. On average, under this setting, the running time for training is $3795.12s$ per epoch and $0.043ms$ for inference per sample.

% Our model is trained based on the following values: batch size is $16384$, learning rate is $0.01$, embedding size and GRU hidden size are both $64$， hidden_dim is 32 and num_gru_layers=2;
% epoch=5;

\begin{table} [t]
\setlength{\abovecaptionskip}{0.3cm}
\centering
 \small
\caption{ The parameter settings  }
\begin{tabular}{llrrrr}
%\hline
%\toprule[1pt]
\toprule
\multirow{1}{*}{ name } & \multirow{1}{*}{value}    \\          

\midrule

    Training  batch size     &     $16384$    \\
     Learning rate      &     $0.01$     \\
      The top-level embedding size       &   64       \\
        GRU hidden size      &    64     \\
          The number of views in similarity vector    &    32     \\
           The number of GRU layers      &    2     \\
            Epochs & 8\\
       
\bottomrule
\end{tabular}\label{tbl:param_model}
% \label{table_2}
\vspace{-2mm}
\end{table}

% number of features;
% All features are standardized by zero-mean and unit-variance normalization before model training.

% time span;
% train-test splits;

% Each dataset is constructed from the interactions among users belonging to a specific country (obscured for privacy reasons). We collect 79 user features spanning four modalities and six pairwise link features, as described in Section 3.1. All features are standardized by zero-mean and unit-variance normalization before model training. In each dataset, the training set comprises timetamped friendships created during a span of 7 contiguous days. Empirically, we find that 7 days suffices to achieve good results (comparable to 1 month), thus significantly more efficient. To evaluate the quality of friendship suggestions, the test set comprises all friend add requests over the subsequent four days. We observe consistent results for different train-test splits across 5 time periods and 2 geographic regions. We use 10\% of the labeled examples for hyper-parameter tuning.

\subsection{Experimental Results}

\subsubsection{Friend suggestion}
We first present the friend suggestion performance of all methods in Table.~\ref{tbl:COMPARSION},  including the Hit rate and NDCG  values on these testing datasets. All these methods are built on same user features, including users' profiles, users' in game behavior sequences and 
pairwise link features?
Generally speaking, the performance shows a similar trend in these datasets that \ourmeth $\geq$ self-Attention $\geq$  GRU $\geq$  MLP. 
We find that MLP always gets the worst performance because it does not consider the temporal dependencies in data, thus fails to capture the dependencies in user sequence features. 
As a comparison, GRU gains decent improvement over MLP since it captures dependencies in sequence by memorizing information. 
These two models both obtain the similarity by directly learning embeddings from raw features. 
This can simplify the similarity between users due to the evolving user preferences. 
Thanks to the modeling similarity evolution,  our method \ourmeth consistently outperforms baseline methods, achieving  around $3\%$- $35\%$ gains over baselines. When removing this similarity evolution as  self-Attention model, we observe noticeable performance decline, indicating the expressiveness of similarity evolution modeling.

\par
% \todo{GraFrank experimental details are missing}
We also compared the performance of our DSEN with GraFrank on the two datasets Section-5 and Section-6, and the results are reported in Table~\ref{tbl:COMPARSION_gnn}. 
As we mentioned before, this is not a fair comparison, so this part is to provide another angle to understand the directions of addressing friend suggestion, using social networks or not. 
Here, we follow the feature usage in the original GraFrank paper, and use the user profile as the input. 
We observe a significant performance gap between ours and GraFrank. 
Apart from being different in the use of features, GNN-based model has the built-in drawback of learning interactions in features, thus it is hard to obtain the similarity between users in our setting. 

% We know that GNNs can learn both local and global structure information of graphs, however they lack the ability to mine high-level feature combinations. 
% Consequently, DSEN and other baselines all outperform the GraFRank on both datasets.

% Consequently, DSEN and other baselines all outperform the GraFRank in all terms of metric on both datasets.

% Note that, in both datasets, we adopt the user's active sequence feature and profile feature when reproducing MLP, GRU, and self-attention, but only the profile features are introduced in GraFRank. Thus, it is not a fair comparison, but some facts can still be reflected from the experimental results. 

% GRU: 没有sim 和attention
% SELF-ATT: 只是没第一层，

\begin{figure} [t]
    \centering
    % \subfigure  % [Training loss]
    % {
    %     \includegraphics[width=0.21\textwidth]{images/train_loss.eps}
    % }\hspace{-0.1in}
    %     \subfigure % [Validation AUC ]
    % {
    %     \includegraphics[width=0.21\textwidth]{images/auc_val.eps} 
    % }\hspace{-0.1in} 
    \subfigure[][Training loss]
 {
  \includegraphics[width = 0.21\textwidth]{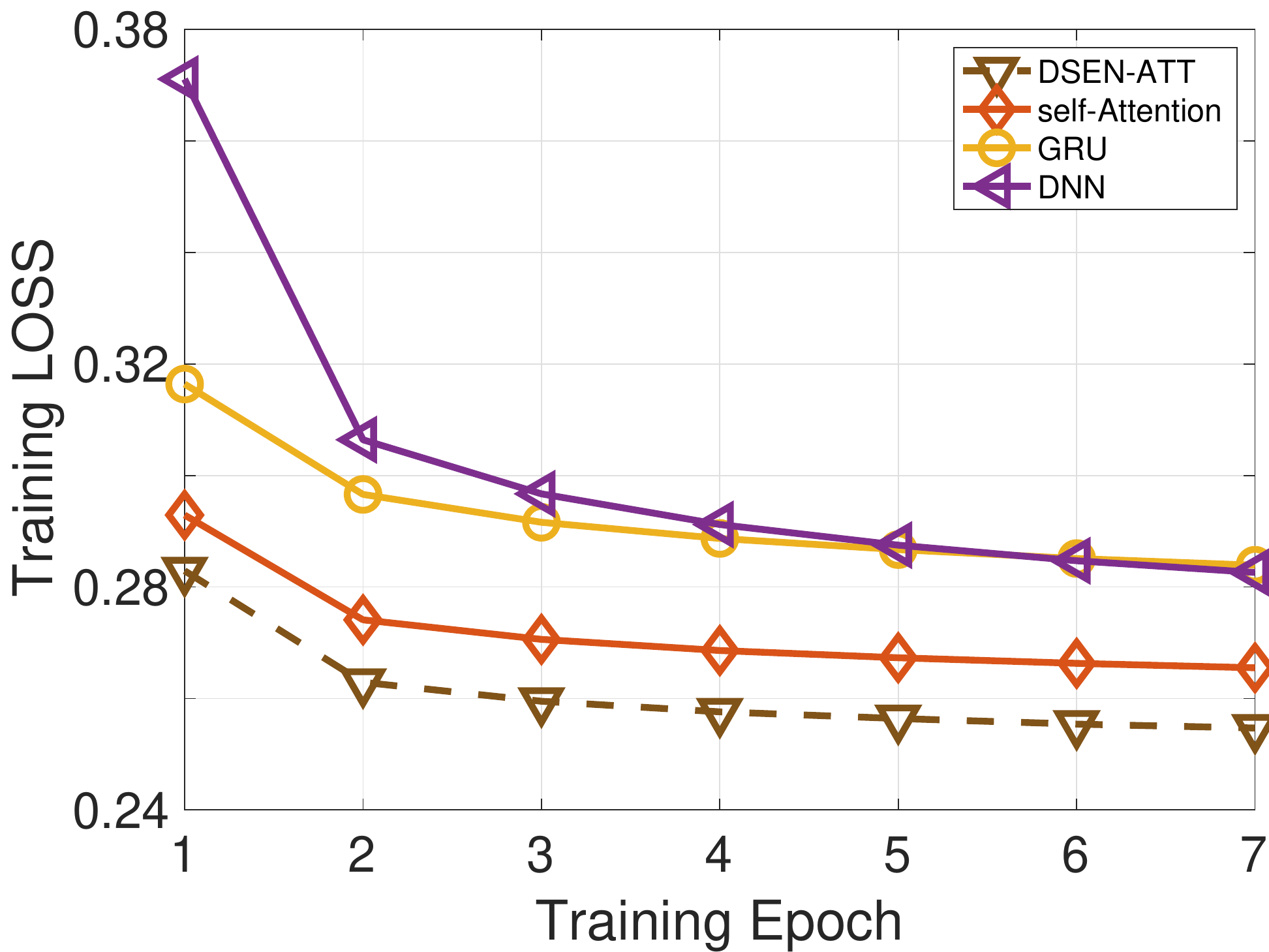}
   \label{fig:var_burstsize}
 }
 \hspace{-0.1in}
 \subfigure[][Validation AUC ]
 {
  \includegraphics[trim={0cm 0cm 0cm 0cm},clip,width=0.21\textwidth]{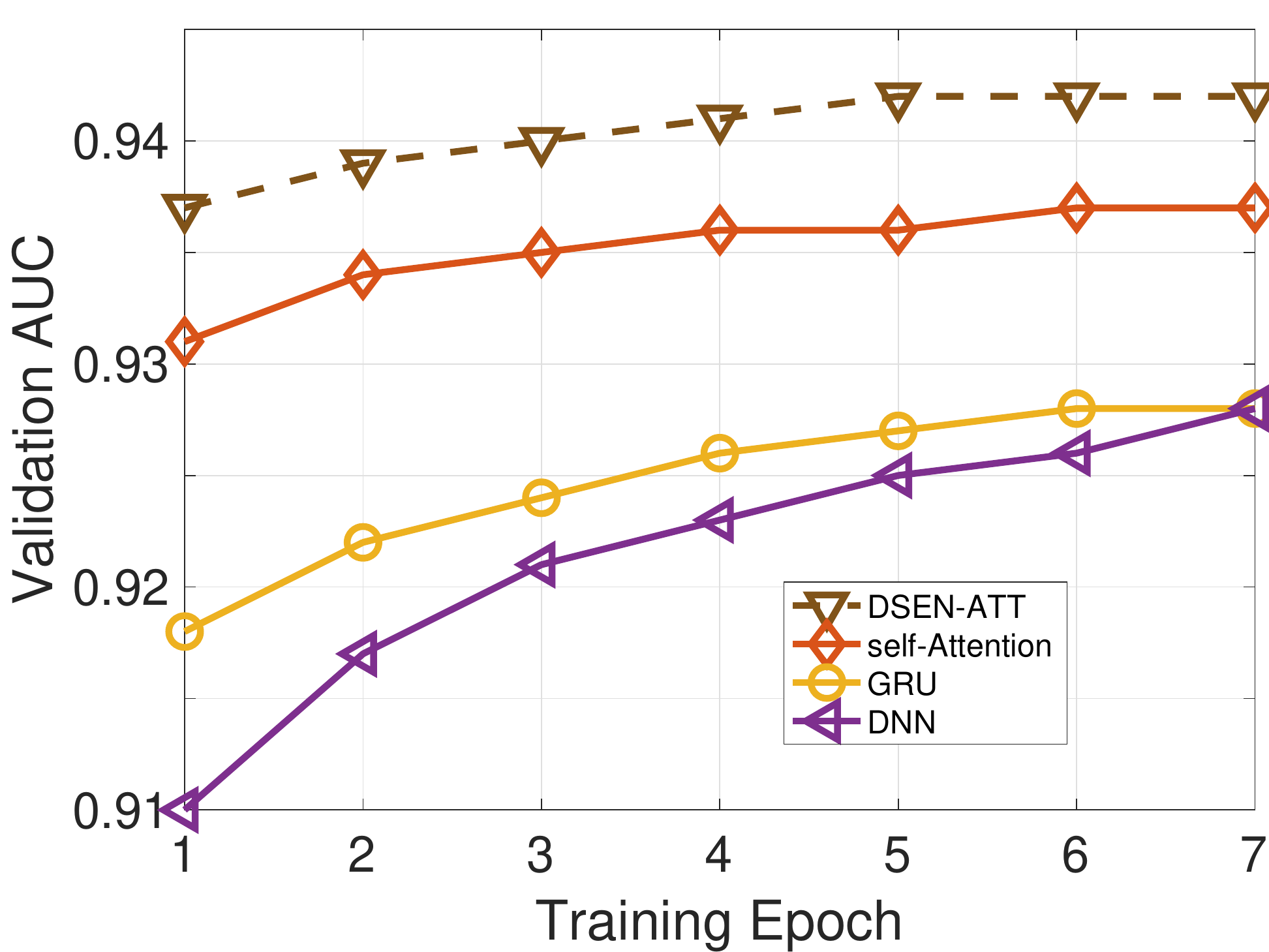} 
  \label{fig:var_global}
  }
    
    \caption{
    %\footnotesize
    The analysis of  training loss and validation AUC: \ourmeth converges faster than other baselines and reaches a better optimization minima w.r.t AUC. }
    \label{fig:loss_auc}
\end{figure}

% \begin{figure} [!t]
%     \begin{minipage}[b]{0.3\linewidth}
%      \captionsetup{font={small}}
% 		\centering
% 		\includegraphics[scale=0.22]{images/train_loss.eps}
% % 		\centerline{\small{}}
% 	\end{minipage}
%     \vspace{.003in}
%     \hspace{.08in}
% 	\begin{minipage}[b]{0.3\linewidth}
%         \captionsetup{font={small}}

% 		\centering
% 		\includegraphics[scale=0.22]{images/auc_val.eps}
% % 		\centerline{\small{(a) HIT@K}}
% 	\end{minipage}
% \end{figure}

\subsubsection{Ablation study}
Now, we present an ablation study to analyze the architectural choice of the model. 
As our goal is to capture the evolution of similarity, LSTM is one immendiate choice and the other is an attention model such as TransFormers~\cite{2017Attention}. 
For comparison, we now implement a version of the TransFormers model,, called as \ourmeth-ATT.
The results of both methods are presented in Table ~\ref{tbl:COMPARSION_ablation}. 
Interestingly, we find that DSEN-ATT performs much worse than DSEN. 
One possible explanation is that the similarity depends more on the most up-to-date activities as opposed to the entire global sequence, so that as the global capture ability becomes a setback of Transformer.
% Because the similarity relation between user-pairs has obvious sequence dependency over time,
LSTMs can capture this contextual relationship more accurately, especially when the sequence similarity is short, which is generally the case in our setting. 
% \todo{add more exp or more analyze}
% . The TransFormer tends to capture global correlations compared to LSTM, and its ability to model position information is weak. Hence, when sequence similarity is short, it may not model sequence evolution well. 

% \begin{table} [t]
% \setlength{\abovecaptionskip}{0.3cm}
% \centering
%  \small
% \caption{We perform ablation analysis on two datasets. Attention-based variational model has inferior performance than the default one. }
% \begin{tabular}{llrrrrrrr}
% %\hline
% %\toprule[1pt]
% \toprule
% \multirow{2}{*}{Dataset} & \multirow{2}{*}{models} & \multicolumn{2}{c}{K=10}         &  & \multicolumn{2}{c}{K=100}                       \\ \cline{3-6}
%                          &                         & HIT@K           & NDCG@K                     & HIT@K           & NDCG@K              \\ %\hline

% \midrule

%                          & DSEN-ATT         &0.1767    &0.0712    & 0.3688 &0.1062     \\
%   Section-2  
%                       & \textbf{DSEN}              & \textbf{0.1914 }      &     \textbf{0.1167   }     & \textbf{0.4050}  & \textbf{0.1600 }            \\
                       
% \midrule

%                          & DSEN-ATT          &0.1633     &0.0675             & 0.3370  & 0.0991     \\
%   Section-3  
%                       & \textbf{DSEN}              &\textbf{0.1846 }      &     \textbf{0.1159}    & \textbf{0.3679}  & \textbf{0.1531}            \\

% \bottomrule
% \end{tabular}\label{tbl:COMPARSION_ablation}
% % \label{table_2}
% \vspace{-2mm}
% \end{table}

\subsubsection{Hyperparameter analysis}

In this subsection, we aim to analyze the effect of a few important hyperparameter of the \ourmeth, including the embedding size of prediction layer, GRU output dimensions, and the number of views in similarity computation.
The results of different values are presented at 
Figure.~\ref{fig_ndcg} and Figure.\ref{fig_hit}, showing Hit rate and NDCG, respectively. \\

\noindent \textit{Effects of the embedding size of prediction layer:} We first analyze the impact of varying the dimension of the top-level embedding in the prediction layer, and report the results in the first column of Figure.~\ref{fig_hit} and Figure.~\ref{fig_ndcg}. 
We observe that with an increase in the dimension of the prediction layer, it promotes performance in HIT@K and NDCG@K. 
These dimensions determine the expressive capability of the model, but an over-sized dimension will significantly increase the learning cost and cause over-fitting in the test data.
So we observe a diminishing returns when the dimension is greater than 64.

\noindent \textit{Effects of GRU  dimensions:} 
We now investigate the influence of the value of GRU dimensions on the behavior sequence extraction layer. 
In the second column in Figure.~\ref{fig_hit} and Figure.~\ref{fig_ndcg}, we present the performance comparison w.r.t the GRU dimension.
Generally, with the increase of GRU dimension, the performance of small dimensions shows significant increase while that of large dimensions has little change. This parameter affect the amount of information that characterize the user's interest, but a too large dimensions may introduce the noise during feature interaction.

\noindent \textit{ Effects of the number of views in similarity:} 
Recall that we developed a multi-view model for similarity computation.
The key hyperparameter is the number of similarity matrices as in Eq.~\ref{VIEW_FUNCTION}. 
This adds a large level of flexibility and non-linearity to the similarity measurement. 
So, we vary this number to understand its effect on the performance. 
From the results, we can observe that large  $\bf k$  tends to obtain better performance, about 1\%-2\% improvement when varying  $\bf k$  from $4$ to $32$.

\subsubsection{Model Training Analysis.}

We now analyze the training of the model by examining the training loss and AUC performance in Figure.~\ref{fig:loss_auc} (a) and (b), respectively.
%From the curve given in figure,
As expected, 
experiments on the train loss and AUC both demonstrate that DSEN has better training efficiency. 
DSEN shows faster convergence over baselines which take about $3$ epochs to reach its optimal state. 
The empirical evidence shows a flat loss landscape that leads to better generalization. 
For this experiment, we observe a similar scene of \ourmeth, achieving the lowest training loss and having results on the test datasets. % \vsa

\section{Conclusion }

We study the friend suggestion problem in large scale online game scenarios, which appears to be the first work on this setting. 
We formulate this as a similarity measurement problem, motivated by our empirical experience, and address it via the proposed similarity evolution model. 
To understand user preference more accurately, we employ both long-term features and short-term features of players in games, which helps us determine the user's current intentions. 
We develop a deep neural framework that tackles the aforementioned features, and learns the preference evolution process from a proposed similarity measurement model. 
We run our experiments on several million user datasets from the a major online game company in the world, and our proposed approach consistently outperforms the competing baselines.

{%\footnotesize
\bibliographystyle{abbrv}
\bibliography{references}
}
\end{document}